\documentclass[backref, breaklinks,colorlinks,onecolumn]{aastex701}
\hypersetup{linkcolor=blue,citecolor=blue,filecolor=cyan,urlcolor=magenta}
\usepackage{enumitem}
\usepackage{booktabs}
\usepackage[fleqn]{amsmath}
\usepackage{amssymb}
\usepackage{xcolor}

\shorttitle{Fermi Pulsars: FP Boundaries \& Visibility}
\shortauthors{Kalapotharakos et al.} %%
\newcommand{\g}{\gamma}

\newcommand{\ec}{\epsilon_{\rm c}}

\newcommand{\ed}{\dot{\mathcal{E}}}

\newcommand{\gL}{\gamma_{\rm L}}
\newcommand{\Rc}{R_{\rm c}}
\newcommand{\BLC}{B_{\rm LC}}
\newcommand{\RLC}{R_{\rm LC}}

\begin{document}
\tighten

\title{A Comprehensive Interpretation of Fermi-LAT Pulsars: Fundamental-Plane Death Border, Visibility Thresholds, and GeV-TeV Unification}

\correspondingauthor{Constantinos Kalapotharakos}

\author[0000-0003-1080-5286]{Constantinos Kalapotharakos}
\affiliation{Astrophysics Science Division, NASA Goddard Space Flight Center, Greenbelt, MD 20771, USA}
\email[show]{constantinos.kalapotharakos@nasa.gov}
\email[show]{ckalapotharakos@gmail.com}

\author[0000-0002-9249-0515]{Zorawar Wadiasingh}
\affiliation{Department of Astronomy, University of Maryland, College Park, College Park, MD 20742, USA} \affiliation{Astrophysics Science Division, NASA Goddard Space Flight Center, Greenbelt, MD 20771, USA} \affiliation{Center for Research and Exploration in Space Science and Technology, NASA GSFC, Greenbelt, MD 20771, USA}
\email{zwadiasingh@gmail.com}

\author[0000-0001-6119-859X]{Alice K. Harding}
\affiliation{Theoretical Division, Los Alamos National Laboratory, Los Alamos, NM 87545, USA}
\email{ahardingx@yahoo.com}
%\collaboration{1}{(AAS Journals Data Scientists collaboration)}

\author[0000-0002-7435-7809]{Demosthenes Kazanas} \affiliation{Astrophysics Science Division, NASA Goddard Space Flight Center, Greenbelt, MD 20771, USA} 
\email{demos.kazanas@nasa.gov}
%\nocollaboration{1}

\author[0009-0001-8783-1004]{Dimitrios Skiathas}
\affiliation{Southeastern Universities Research Association, Washington, DC 20005, USA}
\affiliation{Astrophysics Science Division, NASA Goddard Space Flight Center, Greenbelt, MD 20771, USA}
\affiliation{Center for Research and Exploration in Space Science and Technology, NASA GSFC, Greenbelt, MD 20771, USA}
\affiliation{Department of Physics, University of Patras, Rio, 26504, Greece}
\email{dimitrios.skiathas@nasa.gov}

%\nocollaboration{1}
%\collaboration{1}{(LaTeX collaboration)}



\begin{abstract}
We present a framework that links equatorial-current-sheet (ECS) physics to catalog-level, phase-averaged gamma-ray pulsar properties. Guided by analytic scalings and particle-in-cell (PIC) simulations, we show that the pulsar ``Fundamental Plane'' (relating gamma-ray luminosity, spectral cutoff energy, spin-down power $\dot{\cal{E}}$, and surface magnetic field) is bounded by two regimes: a radiation-reaction-limited branch and a potential-drop-limited branch. Their intersection defines a transition in $\dot{\cal{E}}$ that maps to a gamma-ray visibility threshold on the $P$--$\dot{P}$ diagram, above which detectability is set by distance and beaming, and below which both cutoff energy and efficiency decline rapidly. Placing ATNF pulsars and McGill magnetars onto these planes reproduces the observed Fermi occupancy, with millisecond pulsars (MSPs) on the observable side, young pulsars (YPs) straddling the threshold, and magnetars clustering at or just below it. At higher $\dot{\cal{E}}$, both MSPs and YPs depart from the maximal radiation-reaction-limited envelope at similar cutoff energies, suggesting that enhanced pair creation screens the accelerating electric field in the ECS. We interpret this behavior with a compactness-based criterion for optically thin $\gamma\gamma$ pair feedback in or near the ECS and briefly note an extension to $\gamma\gamma\rightarrow\mu^\pm$ that could yield pulsed multi-TeV neutrinos in the most energetic systems. The framework predicts a MeV-bright, GeV-faint corridor below Fermi sensitivity, a target for next-generation MeV missions. Finally, motivated by the recent HESSII detection of pulsed multi-TeV emission from Vela, we use PIC particle distributions with a seed-photon model to reproduce a multi-TeV inverse-Compton component alongside the GeV curvature emission, supporting a unified ECS-based GeV-TeV origin.
\end{abstract}

\keywords{pulsars: general — stars: neutron — gamma rays: stars — radiation mechanisms: non-thermal — magnetars}


\defcitealias{2013ApJS..208...17A}{2PC}

\defcitealias{2020ApJS..247...33A}{4FGL}
\defcitealias{2022arXiv220111184F}{4FGL-DR3}
\defcitealias{2005AJ....129.1993M}{ATNF}
\defcitealias{2019ApJ...883L...4K}{K19}
\defcitealias{2018ApJ...857...44K}{K18}
\defcitealias{Smith23}{3PC}

\def\lambar{\lambda\llap {--}}

\section{Introduction} 
\label{sec:intro}

The Fermi Large Area Telescope (LAT) has revolutionized our understanding of pulsars as powerful gamma-ray emitters. Since its launch, LAT has uncovered a rich and diverse population of gamma-ray pulsars, including young, energetic pulsars (YPs), and older recycled pulsars (the millisecond pulsars, MSPs), including radio-quiet pulsars. Through precise measurements of their spectra and pulse profiles, LAT observations have shown that pulsed GeV gamma-ray emission is a common feature among energetic rotation-powered pulsars, although the observed sample is shaped by sensitivity and selection effects (e.g., beaming geometry). The energetic output of these pulsars peaks in the LAT band (photon energies of order GeV), providing the most direct constraints on the structure of their magnetospheres, the locations of particle acceleration, and emission models. Collectively, LAT pulsars serve as a powerful probe of magnetospheric physics across a wide range of evolutionary stages and physical parameters.

In parallel with these observational breakthroughs, substantial theoretical progress has transformed our understanding of pulsar magnetospheres. A foundational step was the development of global solutions under the force-free (FF) approximation, where the plasma is sufficiently dense to enforce the ideal conditions $\vec{E} \cdot \vec{B} = 0$ and $E < B$ throughout the magnetosphere for local electric and magnetic fields $E$ and $B$. These FF models revealed the robust formation of a large-scale \textit{equatorial current sheet} (ECS) that emerges at and beyond the light cylinder\footnote{The light cylinder is the cylindrical surface where the corotation speed equals the speed of light, defined by $R_{\rm LC} = c / \Omega$, with $\Omega$ the stellar angular velocity.} \citep{1999ApJ...511..351C,2006MNRAS.368.1055T,2006ApJ...648L..51S,2009A&A...496..495K,2012MNRAS.420.2793K}. In this inherently dissipative structure, magnetic reconnection and particle acceleration can naturally take place \citep{1996A&A...311..172L,2002A&A...388L..29K,2010ApJ...715.1282B,2010MNRAS.404..767C,2011MNRAS.412.1870P}.

To incorporate magnetospheric dissipation and its observational implications, the FF framework was extended into the so-called \textit{FIDO} model (Force-Free Inside, Dissipative Outside), introduced by \citet{2014ApJ...793...97K} \citep[see also][]{2015ApJ...804...84B,2017ApJ...842...80K}. This model introduced a finite plasma conductivity zone outside the light cylinder, enabling particle acceleration and tunable energy dissipation. It was shown to reproduce the general characteristics of LAT gamma-ray light curves and, more importantly, provided a natural setting for the location of the high-energy emission within the ECS. Importantly, \citet{2014ApJ...793...97K} found that the effective ECS emissivity is not uniform but instead concentrated in specific regions (closer to the rotational equator) along the sheet, which plays a key role in setting the observed pulse morphologies. Notably, by assuming that the radio emission originates near the magnetic poles at low altitudes, \citet{2014ApJ...793...97K} successfully reproduced the observed correlation between radio phase lag $\delta$ and gamma-ray pulse profile peak separation $\Delta$ for LAT YPs. This phase agreement localizes the gamma-ray emission to the ECS outside the light cylinder and, together with the resulting non-uniform ECS emissivity, provides a strong geometric constraint on the emission zones.

Further refinement of pulsar emission modeling came through kinetic, three-dimensional global Particle-in-Cell (PIC) simulations \citep[e.g.,][]{2014ApJ...785L..33P,2015ApJ...801L..19P,2016MNRAS.457.2401C,2018ApJ...858...81B,2018ApJ...857...44K,2018ApJ...855...94P,2023ApJ...954..204K}, which self-consistently evolve both the electromagnetic fields and the particle distributions across the entire magnetosphere\footnote{These global PIC models, however, do not capture realistic field strengths, self-consistent relativistic plasma injection from polar-cap pair cascades, or the actual high Lorentz factors achieved near the ECS, making rescaling or extrapolation schemes necessary.}. These simulations suggest that the dominant high-energy emission originates in the ECS, where particles are accelerated to the highest Lorentz factors and emit either curvature or synchrotron radiation, primarily beyond the light cylinder.

A key regulating factor in these global PIC simulations is the rate of pair injection in the so-called \textit{separatrix zone} (SZ), a narrow region near the boundary separating open and closed field lines, where plasma is presumably supplied to the broader ECS by unresolved magnetospheric cascades. The SZ controls the level of local dissipation and, consequently, determines both the gamma-ray luminosity and the spectral cutoff energy of the pulsar emission \citep[][see also \citealt{2017ApJ...842...80K}]{2023ApJ...954..204K}.

In parallel, a deeper yet relatively simple theoretical insight led to the identification of a striking empirical correlation now known as the \textit{Fundamental Plane} (FP) of gamma-ray pulsars using phase-averaged observables. Using a sample of 88 pulsars from the Second  {Fermi} Pulsar Catalog \citep{2013ApJS..208...17A}, \citet{2019ApJ...883L...4K} showed that the gamma-ray luminosity $L_\gamma$ scales with the measured spectral cutoff energy $\epsilon_{\text{c}}$, the surface magnetic field $B_\star$, and the spin-down power $\dot{\cal E}$ as
\begin{equation}
    L_\gamma \propto \epsilon_{\text{c}}^{1.18 \pm 0.24} \, B_\star^{0.17 \pm 0.05} \, \dot{\cal E}^{0.41 \pm 0.08},
    \label{eq:FP_2PC}
\end{equation}
a relation that was interpreted in that work as consistent with curvature radiation in the ECS beyond the light cylinder. In this regime, the relevant radius of curvature is dimensionally set by length scales associated with the light cylinder, thereby coupling the global magnetospheric structure geometry to the radiative process.

This FP relation was later refined in \citet{2022ApJ...934...65K} using a larger pulsar sample from the 4FGL-DR3 catalog, yielding
\begin{equation}
    L_\gamma = 10^{14.3\pm1.3}\epsilon_{\text{c}}^{1.39 \pm 0.17} \, B_\star^{0.12 \pm 0.03} \, \dot{\cal E}^{0.39 \pm 0.05},
    \label{eq:FP_4FGL}
\end{equation}
where $\epsilon_{c}$ is measured in MeV, $B_{\star}$ is measured in G, and $\dot{\mathcal{E}}$ and $L_{\gamma}$ are measured in $\rm erg\,s^{-1}$ and was further corroborated by PIC simulations in \citet{2023ApJ...954..204K}, where the quantities extracted directly from the particle distributions and radiation calculations, rescaled to realistic pulsars, reproduced both the theoretical FP scaling and the observed trends in the LAT data.

A complementary validation of the FP was provided by \citet{2025ApJ...985...58A}, who analyzed pulsars from the third LAT pulsar catalog \citep[][3PC, hereafter]{Smith23} and used machine-learning techniques on the radio-loud population to estimate distances for radio-quiet gamma-ray pulsars. Their study confirmed that the FP is favored as a gamma-ray pulsar luminosity function (for sources with significant spectral cutoffs), effectively ruling out synchrotron emission as the dominant process in the LAT band.

Taken together, these investigations consolidate the FP as a physically grounded and observationally verified framework, strengthened by numerical simulations. They strongly support the interpretation that the high-energy emission in LAT pulsars originates from curvature radiation in the ECS, with the key phase-averaged observables and pulsar parameters ($L_\gamma$, $\epsilon_{\text{c}}$, $\dot{\cal E}$, and $B_\star$) linked by underlying physics rather than selection effects.

Additional insight can be drawn from the observed pulsar population, beyond the LAT detections. While the FP is applicable across a wide parameter space, the observed gamma-ray pulsars occupy only a limited portion of it. Clearly, the FP relation breaks down when ${\cal \dot{E}} < L_\gamma$. Because $d\log L_\gamma/d\log{\cal \dot{E}}< 1 $ in Eq.~\ref{eq:FP_2PC}--\ref{eq:FP_4FGL}, the implied efficiency $L_\gamma/\dot{\mathcal{E}}$ increases toward low $\dot{\mathcal{E}}$; an unbroken extrapolation would eventually yield $L_\gamma>\dot{\mathcal{E}}$, so the FP relation must turn below a characteristic $\dot{\mathcal{E}}$ (set by the normalization and other parameters)\footnote{These efficiency arguments, based on the voltage-like scaling of $L_\gamma$, imply a lower bound ${\cal \dot{E}} \sim {\rm few \times} 10^{32}$~erg/s below which bright gamma-ray emission is untenable; this was first noted by \cite{1996A&AS..120C..49A} for EGRET pulsars.}. 

But what is the form of $L_\gamma$ at low $\cal \dot{E}$? For higher $\cal \dot{E}$, simulations suggest that high-energy emission could be possible in areas of the FP where no pulsars have yet been detected, which suggests additional physics not yet captured by these models. This raises key questions: Why do gamma-ray pulsars populate only a specific region of the FP? Does this reflect observational biases, or does it point to intrinsic physical boundaries (such as gamma-ray death lines or death valleys) defined either in terms of spin-down power $\dot{\cal E}$ or more complex conditions within the FP itself? In this work, we aim to explore these questions and characterize the physical constraints that determine the visibility and viability of pulsed gamma-ray emission across neutron stars.

In this work, we investigate the physical origins of the FP's regime of applicability. We argue that the confinement of observed pulsars to a limited region arises from the presence of a gamma-ray \textit{death border}. By this we mean the upper envelope in FP space defined jointly by the \textit{maximal radiation-reaction-limited} (RRLmax) and \textit{potential-drop-limited} (PDL) branches. This boundary is not a single line but a finite-width band: for a given $\dot{\mathcal E}$, different combinations of $P$ and $B_\star$ yield slightly different theoretical envelopes, which collectively form the practical ``death border.'' The location and shape of this death border are set in part by the evolution of the maximum achievable spectral cutoff energies and by the transition from the RRLmax regime to the PDL regime, where acceleration is governed by the available rotationally driven voltage. These constraints influence the detectability of pulsars, particularly at lower $\dot{\cal E}$ values, and lead us to examine the existence of an undetected population of MeV-bright pulsars that likely lie below the LAT sensitivity threshold. More broadly, the same competition between radiation-reaction limits and a finite available voltage should also arise in other current-sheet accelerators beyond isolated pulsars, for example, in the dynamically interacting magnetospheres of merging binary neutron stars \citep{2025ApJ...994..131S}.

Together with the distribution of observed gamma-ray pulsars on the FP, our results provide a comprehensive framework for understanding the phase-averaged gamma-ray emission of pulsars, highlighting the role of pair-production efficiency, set by intrinsic pulsar properties such as $B_{\star}$ and rotation period $P$, in regulating which pulsars can sustain bright MeV-GeV emission. At the same time, while the FP and RRL scalings capture most of the Fermi-LAT population, the highest-$\dot{\mathcal E}$ sources do not follow the RRLmax track indefinitely: their spectral cutoff energies, $\epsilon_{\rm c}$, flatten relative to the envelope, indicating additional screening of the accelerating electric field component, $E_{\rm acc}$. This plausibly arises from enhanced pair creation in the ECS (or in other magnetospheric regions that supply pairs to the ECS), which modulates the effective accelerating scale. To explore a physical trigger for this behavior, we introduce a compactness-based criterion for optically thin $\gamma\gamma$ \emph{pair} feedback ($e^{\pm}$ creation) in or near the ECS and discuss its population-level implications. As an extension of the same activation picture, we also outline the possible opening of the $\gamma\gamma\rightarrow \mu^{\pm}$ channel and the corresponding prospects for pulsed multi-TeV neutrinos in the most energetic systems. We also quantify this $e^{\pm}$ pair-regulated deviation and its implications for population visibility and spectra.

Our findings also indicate that distinct visibility thresholds may apply to YPs and MSPs, motivating us to extend our analysis to other classes of neutron stars such as magnetars and transitional high-B pulsars, most of which have thus far eluded GeV gamma-ray detection.
We show that some high-B pulsars lie near the gamma-ray detection threshold, suggesting that rotation-powered ECS emission peaking in the MeV band may occur in certain high-$\cal \dot{E}$ magnetars. This component would coexist with magnetically-powered inner magnetospheric emission from resonant Compton scattering in some magnetars \citep{2007Ap&SS.308..109B,2011ApJ...733...61B,2013ApJ...762...13B,2018ApJ...854...98W,2019BAAS...51c.292W,2025ApJ...991..178H}. Future MeV-sensitive instruments with phase-resolved spectropolarimetric capabilities might be able to disentangle these distinct magnetically and rotationally powered emission components.

Finally, we turn our attention to the recent remarkable detection of pulsed TeV emission from the Vela pulsar phase-aligned with the LAT GeV component \citep{2022tsra.confE..33D,2023NatAs...7.1341H}. To assess whether this component can be understood within the same physical framework, we connect our theoretical and computational models with these observations. Using particle energy distributions from our PIC simulations, combined with a simplified prescription of the target photon field, we investigate how inverse Compton (IC) spectra evolve with the magnetospheric particle population. Our goal is to determine whether both the GeV and TeV emission from Vela can be explained self-consistently through the ECS-residing particle population.

The paper is organized as follows. In Section~\ref{sec:Death Lines on the Fundamental Plane}, we develop the theoretical framework, deriving the RRLmax and PDL branches, the transition $\cal \dot{E}$, and the resulting $\epsilon_{\rm c}$ and $L_\gamma$ scalings (including parameter-dependence through dimensionless scale factors). Section~\ref{sec:PIC Simulations} confronts these scalings with 3D PIC results, projecting models onto the FP, quantifying degeneracies among effective scaling factors, and comparing YP and MSP branches. In Section~\ref{sec:Observational consequences}, we map the framework to observations: we place Australia Telescope National Facility \citep[ATNF;][]{2005AJ....129.1993M}\footnote{\url{https://www.atnf.csiro.au/research/pulsar/psrcat}} pulsars and McGill magnetars \citep{2014ApJS..212....6O}\footnote{\url{http://www.physics.mcgill.ca/~pulsar/magnetar/main.html}} on the projected FP, derive a practical gamma-ray visibility threshold on the $P$-$\dot P$ diagram, and identify the MeV-bright/GeV-faint corridor relevant to next-generation MeV missions.  In Section~\ref{sec:The TeV regime}, we extend to TeV energies, using (rescaled) PIC particle distributions plus a simple seed-photon prescription to explore IC components and their consistency with Fermi-LAT and H.E.S.S.\ observations of Vela. Finally, Section~\ref{sec:Discussion Conclusions} summarizes our main conclusions, synthesizes the results, introduces a compactness-based criterion for in-situ $\gamma\gamma\rightarrow e^{\pm}$ pair creation, contrasts the curvature-ECS and synchrotron viewpoints, and outlines implications, open questions, and (as an exploratory extension) the prospects for pulsed multi-TeV neutrino emission via $\gamma\gamma\!\rightarrow \mu^{\pm}$.

\section{Death border on the FP}\label{sec:Death Lines on the Fundamental Plane}

In this section, we identify and characterize gamma-ray death lines and emission regimes, examining how they arise from basic radiative constraints and how they manifest on the FP.

\subsection{Radiation reaction limit}

The characteristic spectral cutoff energy in the curvature radiation limit, $\ec$, is given by \citep{1975clel.book.....J}
\begin{equation}
    \ec = \frac{3 c \hbar \gL^3}{2 \Rc},
    \label{eq:cutoff energy}
\end{equation}
where $\gL$ is the Lorentz factor of the emitting particle, $c$ the speed of light, $\hbar$ the reduced Planck constant, and $\Rc$ the corresponding radius of curvature. The associated single-particle curvature-radiation energy-loss rate (radiated power) is
\begin{equation}
\mathcal{P}_{\rm curv}=\frac{2 q_e^2 c\, \gL^4}{3 \Rc^2}
    \label{eq:eq:curv_power}
\end{equation}
where $q_e$ is the electron charge.

In the RRL regime of continuous losses, the rate of energy gain from the accelerating electric field balances the radiative energy losses. Using $\dot{\gamma}_{{\rm L}_{\rm loss}}=\mathcal{P}_{\rm curv}/(m_e c^2)$, this balance condition can be written as
\begin{equation}
\begin{split}
    \dot{\gamma}_{{\rm L}_{\rm acc}} &=\dot{\gamma}_{{\rm L}_{\rm loss}}\\
    \frac{q_e c\, \eta_{\BLC} \BLC}{m_e c^2} &= \frac{2 q_e^2 \gL^4}{3 m_e c \Rc^2},
    \label{eq:radiation reaction limit}
\end{split}
\end{equation}
where $m_e$ is the electron mass, $\BLC$ is the magnetic field at the light cylinder, and $\eta_{\BLC}$ is the accelerating electric field, $E_{\rm acc}$ (in units of $\BLC$).

Assuming a radius of curvature $\Rc = \eta_{\RLC} \RLC$ (with $\RLC = Pc / (2\pi)$ the light cylinder radius), and $\BLC = B_{\star} r_{\star}^3 / \RLC^3$ where $B_{\star}$ and $r_{\star}$ are the stellar surface magnetic field and radius respectively, and using the expression for the spin-down power in dipolar force-free (FF) magnetospheres
\begin{equation}
    \ed = \eta_{\alpha}\frac{16\pi^4 B_\star^2 r_\star^6}{c^3 P^4},
    \label{eq:spin down power}
\end{equation}
with $P$ the stellar period and $\eta_{\alpha} \approx 1 + \sin^2\alpha$ modulating the FF spin-down as a function of the magnetic inclination angle $\alpha$ \citep{2006ApJ...648L..51S,2009A&A...496..495K,2012MNRAS.424..605P,2013MNRAS.435L...1T}, ranging from $\approx 1$ (aligned rotator) to $\approx 2$ (orthogonal rotator), one can derive from Eq.~\ref{eq:radiation reaction limit} the Lorentz factor in the RRL regime, ${\gL}_{\rm RRL}$:
\begin{equation}
\begin{split}
    \gamma_{\rm L}^{\rm RRL} &= \left( \frac{3\pi r_{\star}^3 B_{\star}}{q_e c P} \right)^{1/4} \eta_{\BLC}^{1/4} \eta_{\RLC}^{1/2} \\
                  &\approx 5.1 \times 10^7\, \eta_{\BLC}^{1/4} \eta_{\RLC}^{1/2} r_6^{3/4} B_{12}^{1/4} P_{-1}^{-1/4},
    \label{eq:gamma RRX}
\end{split}
\end{equation}
where $B_{12} = B_\star / 10^{12}$\,G, $P_{-1} = P / 0.1$\,s, and $r_6 = r_\star / 10^6$\,cm. Note the weak dependence on $B_{12}$ and $P_{-1}$.

Substituting back into Eq.~\ref{eq:cutoff energy}, the corresponding maximum spectral cutoff energy in the RRL regime is:
\begin{equation}
\begin{split}
    \ec^{\rm RRL} &= \frac{3^{7/4} \hbar c^{9/16}}{2^{7/4} q_e^{3/4} r_\star^{3/8}} \frac{\ed^{7/16}}{B_\star^{1/8}} \eta_{\RLC}^{1/2} \eta_{\BLC}^{3/4} \eta_{\alpha}^{-7/16}\\
        &\approx 10.2\, \eta_{\RLC}^{1/2} \eta_{\BLC}^{3/4} \eta_{\alpha}^{-7/16} \ed_{36}^{7/16} B_{12}^{-1/8} r_6^{-3/8}~{\rm GeV},
    \label{eq:ecut RRX}
\end{split}
\end{equation}
where $\ed_{36} = \ed / 10^{36}\,\mathrm{erg\,s^{-1}}$. 
The dependence of $\ec^{\rm RRL}$ on $\mathcal{\dot{E}}$ and $B_{\star}$ had been demonstrated in \cite{2017ApJ...842...80K} but without the details regarding the scaling factors.

Finally, we note that the generic RRL relations above are valid for any $\eta_{\BLC}$, while the RRLmax regime is defined by setting $\eta_{\BLC}=\eta_{\BLC}^{\max}$, the largest value compatible with a quasi-FF global magnetospheric structure. This choice yields the upper-envelope Lorentz factors, $\gL^{\rm RRLmax}$, and cutoff energies, $\ec^{\rm RRLmax}$, that define the death-border. 

\subsection{Available Potential Drop}

Equation~\ref{eq:ecut RRX} holds under the assumption that particles can reach the Lorentz factor given by Eq.~\ref{eq:gamma RRX}. However, the maximum energy that particles can attain is constrained by the available potential drop they experience and the energy losses they incur along their entire trajectory through the magnetosphere, up to the point where they emit gamma-rays.

Assuming that particles are accelerated by a fraction $\eta_{\rm pc}$ of the total potential drop available from the global open-field-line rotationally-induced potential (here normalized to the polar-cap potential of an aligned, $\alpha=0^{\circ}$, force-free rotator, and encompassing dissipation/acceleration in the outer magnetosphere, e.g., ECS reconnection), the maximum attainable Lorentz factor, provided it remains below the RRL, is given by
\begin{equation}
\begin{split}
    \gL^{\rm PDL} &= \eta_{\rm pc}\frac{q_e B_{\rm LC}R_{\rm LC}}{m_e c^2} \\
                    &=\eta_{\rm pc} \frac{4\pi^2 q_e r_\star^3 B_\star}{m_e c^4 P^2}
    \approx 2.6 \times 10^9\,\eta_{\rm pc} r_6^3 B_{12} P_{-1}^{-2}\;.
    \label{eq:gammaL max}
\end{split}
\end{equation}

Substituting Eq.~\ref{eq:gammaL max} into Eq.~\ref{eq:cutoff energy} yields the corresponding maximum cutoff energy:
\begin{equation}
\begin{split}
    \ec^{\rm PDL} &= \frac{3 \hbar q_e^3}{2 m_e^3 c^{27/4}} \frac{\eta_{\rm pc}^3}{\eta_{\RLC} \eta_{\alpha}^{7/4}} \frac{\ed^{7/4}}{r_\star^{3/2} B_\star^{1/2}} \\
               &\approx 2.8\times10^6~\mathrm{GeV} \cdot r_6^{-3/2} \ed_{36}^{7/4} B_{12}^{-1/2} \eta_{\rm pc}^3  \eta_{\RLC}^{-1} \eta_{\alpha}^{-7/4}.
\label{eq:ecut max}
\end{split}
\end{equation}

While these values appear large at first glance, we will show below that the strong dependence on the spin-down power $\ed$ significantly reduces the expected cutoff energies for pulsars with lower $\ed$.

\subsection{Transition Between Regimes}

The analysis above implies two limiting branches for the maximum attainable particle energy: a curvature-loss RRL branch and a PDL branch such that $\gL\sim \min(\gL^{\rm PDL}, \gL^{\rm RRLmax})$. For the \emph{death-border envelope} we are specifically interested in the transition between the RRLmax realization and the PDL branch. The envelope transition occurs when the two limiting Lorentz factors are equal:
\begin{equation}
\begin{split}
    \gL^{\rm RRLmax} &= \gL^{\rm PDL} \\
    \left( \frac{3\pi r_\star^3 B_\star}{q_e c P} \right)^{1/4} \eta_{\BLC}^{1/4} \eta_{\RLC}^{1/2} 
    &= \eta_{\rm pc} \frac{4\pi^2 q_e r_\star^3 B_\star}{m_e c^4 P^2},
\end{split}
\label{eq:transition condition}
\end{equation}
where, for notational simplicity, we henceforth write $\eta_{B_{\rm LC}}$ to mean $\eta_{B_{\rm LC}}^{\max}$ whenever we refer to the death-border (RRLmax) envelope, unless stated otherwise.

Solving this expression and using Eq.~\ref{eq:spin down power}, we find that the transition occurs at the following spin-down power:
\begin{equation}
\begin{split}
    \ed_{\rm TR} &=\left( \frac{3 m_e^4 c^{39/4}}{2 q_e^5} \right)^{4/7} r_\star^{6/7} B_\star^{2/7} \eta_{\RLC}^{8/7} \eta_{\BLC}^{4/7} \eta_{\rm pc}^{-16/7} \eta_{\alpha} \\
                 &\approx 7.3 \times 10^{31}~\mathrm{erg\,s^{-1}} \cdot r_6^{6/7} B_{12}^{2/7} \eta_{\RLC}^{8/7} \eta_{\BLC}^{4/7} \eta_{\rm pc}^{-16/7} \eta_{\alpha}.
\label{eq:transition spin-down B}
\end{split}
\end{equation}
Unless explicitly stated otherwise, $\dot{\mathcal{E}}_{\rm TR}$ denotes the \emph{envelope} transition computed for RRLmax, i.e., with $\eta_{\BLC}=\eta_{\BLC}^{\max}$.

This result indicates that along the death-border envelope, for $\ed \lesssim \ed_{\rm TR}$ the maximum cutoff energy follows the PDL branch set by the available potential drop, whereas for $\ed \gtrsim \ed_{\rm TR}$ it follows the RRLmax branch (i.e., the upper-envelope RRL scaling). This transition sets a natural power scale and boundary in the emission behavior across the pulsar population.

Using Eq.~\ref{eq:spin down power}, we can alternatively express the transition spin-down power, $\dot{\mathcal{E}}_{\rm TR}$, in terms of the pulsar period $P$
\begin{equation}
\begin{split}
    \ed_{\rm TR} &= \left(\frac{1}{2\pi}\right)^{2/3}\left( \frac{3 m_e^4 c^{21/2}}{2 q_e^5}  \right)^{2/3} P^{2/3} \eta_{\RLC}^{4/3} \eta_{\BLC}^{2/3} \eta_{\rm pc}^{-8/3} \eta_{\alpha}\\
                 &\approx 1.2 \times 10^{32}~\mathrm{erg\,s^{-1}} \cdot P^{2/3} \eta_{\RLC}^{4/3} \eta_{\BLC}^{2/3} \eta_{\rm pc}^{-8/3} \eta_{\alpha},
\label{eq:transition spin-down P}
\end{split}
\end{equation}
which is independent of $r_\star$ and only depends on fundamental constants.

\begin{figure}
    \centering
    \includegraphics[width=0.5\linewidth]{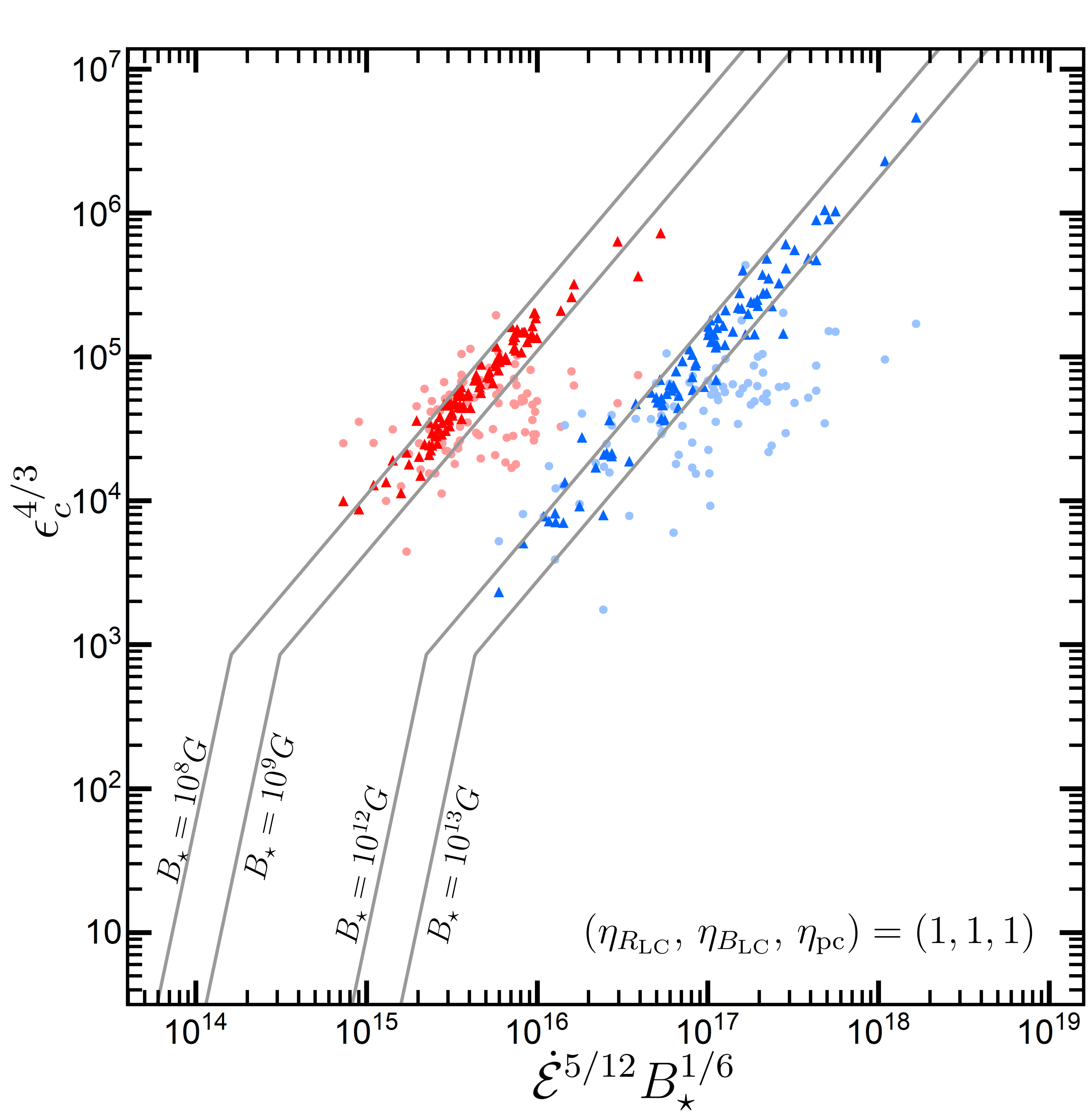}
    \caption{Projected FP of  {Fermi}-detected gamma-ray pulsars. Light-blue filled circles: MSPs; light-red filled circles: YPs, compiled from \citet{2022ApJ...934...65K}. Dark-blue and dark-red triangles show, for the same objects, the locations obtained by adopting the theoretical maximum cutoff energies $\epsilon_{\rm c}$ from Eqs.~\ref{eq:ecut RRX} and \ref{eq:ecut max} with $\eta_{B_{\rm LC}}=\eta_{R_{\rm LC}}=\eta_{\rm pc}=1$, $r_{\star}=10^6$~cm, and $\eta_{\alpha}=3/2$. Gray curves indicate loci of the death lines, i.e., maximum $\epsilon_{\rm c}$ along RRL and potential-drop limited regimes, for constant surface field $B_\star=10^{8},\,10^{9},\,10^{12},\,10^{13}$~G values. The ``knees'' along the curves mark the transition from the maximum RRL to the potential-drop-limited branch (see Eqs.~\ref{eq:transition spin-down B} and \ref{eq:transition spin-down P}). Throughout this and all subsequent FP projections, $\epsilon_{\rm c}$, $\dot{\mathcal E}$, and $B_\star$ are expressed in MeV, ${\rm erg\,s^{-1}}$, and G, respectively. \emph{Note:} For the adopted scaling factors, the theoretical (triangle) locations are effectively deterministic and depend mainly on the measured $\dot{\mathcal E}$ and $B_\star$. By contrast, the catalog-based cutoff proxy values used for the circles inherit the spectral-fit uncertainties and parameter covariances. Individual cutoff error bars are omitted for clarity; for the vast majority of sources, the quoted cutoff uncertainties are $\lesssim 0.3$~dex (often substantially smaller).}
    \label{fig:projectedFP Fermi pulsars and maximum ecut}
\end{figure}

The death lines, i.e., maximum $\epsilon_{\rm c}$ along RRLmax and PDL regimes, described by Eqs.~\ref{eq:ecut RRX} and \ref{eq:ecut max} depend on the magnetic field strength and various scaling factors. In Fig.~\ref{fig:projectedFP Fermi pulsars and maximum ecut}, we show a projection of the FP onto the $\dot{\mathcal{E}}^{5/12} B_\star^{1/6}$---\,$\ec^{4/3}$ plane (in $\log-\log$ scale), based on the FP visualization presented in Fig.~14 of \cite{2023ApJ...954..204K}. The colored circular points show the Fermi-detected pulsars using the $\epsilon_{c1}$ values compiled in \citet{2022ApJ...934...65K} from the 4FGL-DR3 analysis (which encompasses the LAT pulsar sample used in \citetalias{Smith23}).

Light-colored disk-shaped markers indicate the actual positions of YPs (light blue) and MSPs (light red). The darker triangular markers correspond to the $\ec$ values predicted by the theoretical death lines given by Eqs.~\ref{eq:ecut RRX} and \ref{eq:ecut max}, assuming $\eta_{\RLC} = \eta_{\BLC} = \eta_{\rm pc} = r_6 = 1$ and $\eta_{\alpha} = 3/2$\footnote{Throughout this work, we adopt a representative inclination angle of $\alpha = 45^{\circ}$. Unless otherwise stated, we therefore set $\eta_{\alpha} = 3/2$.}. These triangular points are seen to approximately trace the upper envelope of the observed $\ec$ distribution.

We also include in the exact figure theoretical death lines for representative surface magnetic field strengths of $B_\star = 10^8$, $10^9$, $10^{12}$, and $10^{13}$~G, under the same assumptions. These curves illustrate how the maximum achievable cutoff energy varies with both spin-down power and magnetic field.

\begin{figure}
    \centering
    \includegraphics[width=0.75\linewidth]{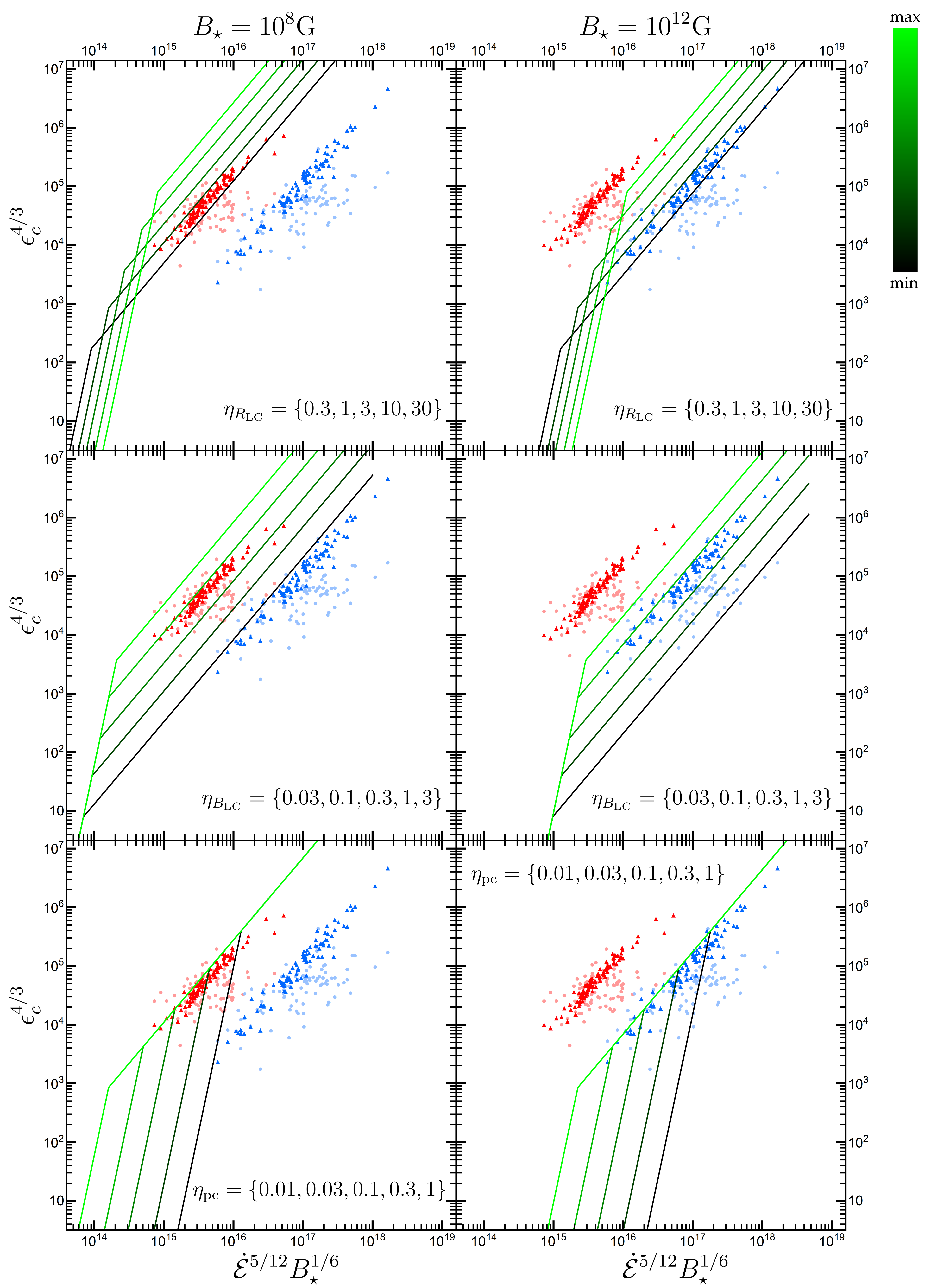}
    \caption{Projected FP for the  {Fermi}-LAT pulsars as in Fig.~\ref{fig:projectedFP Fermi pulsars and maximum ecut}. Each panel overlays families of death lines obtained by varying the indicated scaling factor. Left-hand column: $B_\star=10^{8}$~G (MSP branch). Right-hand column: $B_\star=10^{12}$~G (YP branch). The color scale indicates increasing values of the varied parameter. Except for the varied factor, all other scalings are fixed to unity, with $r_\star=10^6$~cm and $\eta_\alpha=3/2$.}
    \label{fig:projectedFP Fermi pulsars and maximum ecut etaR-etaB-etaPC}
\end{figure}

In the top row of Fig.~\ref{fig:projectedFP Fermi pulsars and maximum ecut etaR-etaB-etaPC}, we illustrate the effect of varying $\eta_{\RLC}$. Each panel shows five curves for $\eta_{\RLC}=0.3,\,1,\,3,\,10,\,30$, assuming $\eta_{\BLC}=\eta_{\rm pc}=r_6=1$. For clarity, we show the plots for MSPs ($B_\star = 10^8$~G) and YPs ($B_\star = 10^{12}$~G) separately, with the left panel corresponding to the former and the right panel to the latter. As $\eta_{\RLC}$ increases, both the maximum cutoff energy $\ec$ in the RRLmax regime and the transition spin-down power $\dot{\mathcal{E}}_{\mathrm{TR}}$ shift to higher values. Consequently, within the PDL regime and at fixed $\dot{\mathcal{E}}$, the maximum $\ec$ decreases with increasing $\eta_{\RLC}$. (Across the knee, the apparent trend can change simply because the operating regime changes; the statements above apply at a fixed regime.)

In the middle row, we vary $\eta_{\BLC}$, plotting five curves for $\eta_{\BLC}=0.03,\,0.1,\,0.3,\,1,\,3$\footnote{Values $\eta_{\BLC}>1$ imply that the ECS and its reconnection layer extend inward of the light cylinder, i.e., the ECS originates interior of $\RLC$, \citep[e.g., see][]{2023ApJ...954..204K}.} with $\eta_{\RLC}=\eta_{\rm pc}=r_6=1$. Again, the left/right panels correspond to MSPs/YPs. Increasing $\eta_{\BLC}$ strengthens the accelerating field, raising the RRLmax $\ec$ and shifting $\dot{\mathcal{E}}_{\mathrm{TR}}$ to higher values. The death line associated with the PDL regime remains fixed.

Finally, in the bottom row, we examine the influence of $\eta_{\rm pc}$ by showing five curves for $\eta_{\rm pc}=0.01,\,0.03,\,0.1,\,0.3,\,1$, with $\eta_{\RLC}=\eta_{\BLC}=r_6=1$. The $\ec^{\rm RRLmax}$ is unaffected, while the transition spin-down power $\dot{\mathcal{E}}_{\mathrm{TR}}$ decreases as $\eta_{\rm pc}$ increases. Thus, the knee shifts to lower $\dot{\mathcal{E}}$ and, correspondingly, to lower $\ec$ at the transition.

The preceding analysis makes it evident that different combinations of scaling factors can, in principle, reproduce nearly identical death-line envelops. In practice, however, the observed behavior in real pulsars likely results from a mix of different particle populations, each characterized by distinct parameter values.

\begin{figure}
    \centering
    \includegraphics[width=0.95\linewidth]{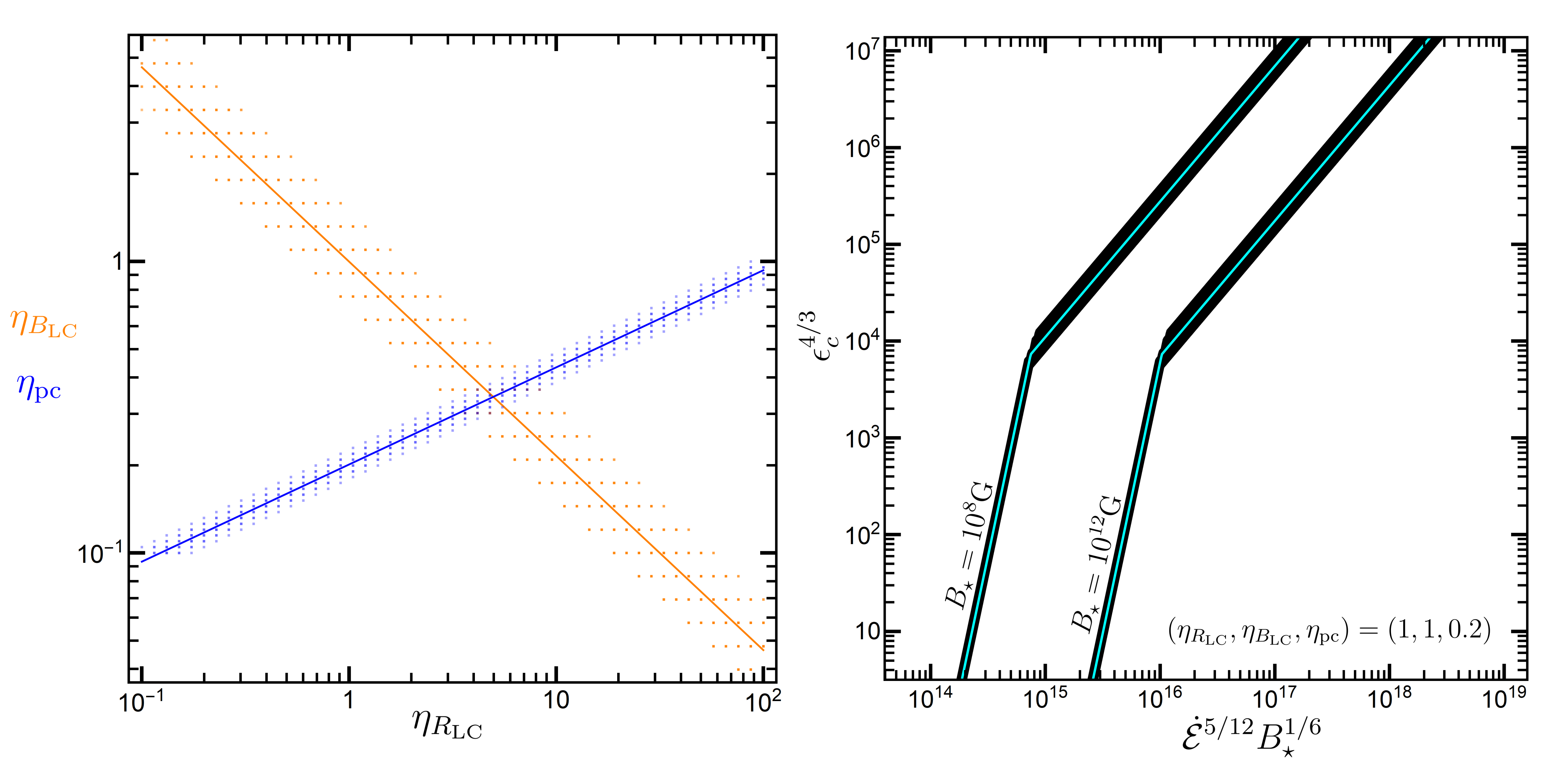}
    \caption{\textbf{Left-hand panel:} Degeneracy curves showing $\eta_{B_{\rm LC}}$ (orange) and $\eta_{\rm pc}$ (blue) as functions of $\eta_{R_{\rm LC}}$ that reproduce the same death-line envelope as the reference choice $(\eta_{R_{\rm LC}},\,\eta_{B_{\rm LC}},\,\eta_{\rm pc})=(1,1,0.2)$, which matches well the trends in the observed pulsars and the PIC simulations (see Fig.~\ref{fig:FP projected Fermi PIC Theory - moving average trend} below). \textbf{Right-hand panel:} The corresponding reference-case envelope is shown by the cyan curves for $\eta_\alpha=3/2$, $r_\star=10^6$~cm, and the indicated $B_\star$ values. The orange and blue point distributions in the left-hand panel map onto the black bands in the right-hand panel, illustrating the range of $(\eta_{R_{\rm LC}},\,\eta_{B_{\rm LC}},\,\eta_{\rm pc})$ combinations that are observationally degenerate around the reference solution.}
    \label{fig:projectedFP Fermi pulsars and maximum ecut degeneracies}
\end{figure}

More specifically, within a given population (MSPs or YPs; i.e., fixed representative $B_\star$) and for fixed $\eta_\alpha$ and $r_\star$, a single death-line envelope implies a fixed transition spin-down power, $\dot{\mathcal{E}}_{\rm TR}$. From Eq.~\ref{eq:transition spin-down B}, this fixes the relation $\eta_{\rm pc}=H_1(\dot{\mathcal{E}}_{\rm TR})\,\eta_{R_{\rm LC}}^{1/2}\,\eta_{B_{\rm LC}}^{1/4}$, where $H_1$ collects all constants that depend only on $\dot{\mathcal{E}}_{\rm TR}$ (for the adopted $\eta_\alpha$ and $r_\star$). In addition, the normalization of the RRL segment is set by the product $H_2\equiv \eta_{B_{\rm LC}}\,\eta_{R_{\rm LC}}^{2/3}$. Thus, matching a specific RRL segment implies $\eta_{B_{\rm LC}}=H_2\,\eta_{R_{\rm LC}}^{-2/3}$, which reduces the above relation to $\eta_{\rm pc}=H_1(\dot{\mathcal{E}}_{\rm TR})\,H_2^{1/4}\,\eta_{R_{\rm LC}}^{1/3}$. In the left-hand panel of Fig.~\ref{fig:projectedFP Fermi pulsars and maximum ecut degeneracies}, we plot (in $\log-\log$ space) the corresponding degeneracy curves $\eta_{B_{\rm LC}}(\eta_{R_{\rm LC}})$ (orange) and $\eta_{\rm pc}(\eta_{R_{\rm LC}})$ (blue) that reproduce the reference case $(\eta_{R_{\rm LC}},\eta_{B_{\rm LC}},\eta_{\rm pc})=(1,1,0.2)$ assuming $r_6=1$, $\eta_{\alpha}=3/2$ and the indicated $B_\star$ values (cyan curves on the projected FP in the right-hand panel). This reference case appears to match well the trends of the currently observed data and the PIC simulations (see Fig.~\ref{fig:FP projected Fermi PIC Theory - moving average trend} below). The orange and blue point distributions in the left-hand panel show the sets of $(\eta_{R_{\rm LC}},\eta_{B_{\rm LC}})$ and $(\eta_{R_{\rm LC}},\eta_{\rm pc})$, respectively, that map into the black bands around this reference solution in the right-hand panel. While the intrinsic degeneracy limits the ability to constrain each scale factor independently, the observed pulsar distribution and the simulation trends nonetheless exclude combinations that do not reproduce the empirical envelope and projected-FP behavior.

\subsection{Gamma-ray Efficiency}

One of the key assumptions underlying the FP derivation is that the number of emitting particles in the dissipative region scales (for all pulsars) as the Goldreich-Julian (GJ) number density at the light cylinder, $n_{\rm GJ-LC}$, multiplied by the dissipative volume, $V_d$. Assuming $V_d \propto R_{\rm LC}^3$ and $n_{\rm GJ-LC} \propto n_{\rm GJ\star} R_{\rm LC}^{-3}$, where $n_{\rm GJ\star}$ is the GJ density at the stellar surface, $n_{\rm GJ-LC}\times V_d$ becomes independent of the light cylinder radius and depends solely on $n_{\rm GJ\star}$.

The notable agreement between this theoretical FP relation, the one derived from observational data, and the one obtained from PIC simulations~\citep{2023ApJ...954..204K}, supports the validity of this scaling assumption. In practical terms, this implies that the number of emitting particles, when expressed in units of the GJ density at the stellar surface, remains roughly constant across the pulsar population. That is, the product $M_{\rm em} \, f_V$ does not substantially vary, where $M_{\rm em}$ is a dimensionless emission multiplicity factor and $f_{V}$ the dissipative dimensionless volume measured in $R_{\rm LC}^3$ units. This steadiness implies that either both $M_{\rm em}$ and $f_V$ are roughly invariant across pulsars, or that variations in one are compensated by inverse variations in the other. 

Given the GJ density at the stellar surface for a quasi-aligned rotator, $n_{\rm GJ\star}=B_\star/(c\, q_e P)$, together with Eqs.~\ref{eq:spin down power} and \ref{eq:cutoff energy}, and following the scaling argument of \citet{2019ApJ...883L...4K}, we take the per-particle gamma-ray power, $L_{\gamma,1} = \mathcal{P}_{\rm curv}$ for curvature losses (Eq.~\ref{eq:eq:curv_power}). If the instantaneous number of emitting particles in the dissipation zone scales as\footnote{Equivalently using the (primary) GJ particle flux from both polar caps, $q_e \dot{N}_{\rm GJ} \sim \rho_{\rm GJ} A_{\rm pc} c$ which yields $\dot{N}_{\rm GJ} \sim 4 \pi^2 B_\star r_\star^3/(q_e c P^2)$. If the emitting path length at the light cylinder is $\sim \RLC$ and so the characteristic residence time $\sim R_{\rm LC}/c$, $N_{\rm em} \sim \dot{N}_{\rm GJ} R_{\rm LC}/c \sim B_{\rm LC} R_{\rm LC}^2/q_e \sim n_{\rm GJ\star}\,r_\star^3$.} 
$N_{\rm em}=M_{\rm em}\,n_{\rm GJ-LC}\,V_d=M_{\rm em}\,f_V\,n_{\rm GJ\star}\,r_\star^3$, then the total luminosity is $L_\gamma=N_{\rm em}\,L_{\gamma,1}$. Defining the emitting multiplicity-volume constant $C_{\rm MV}\equiv M_{\rm em}f_V$, this can be written compactly as $L_\gamma=C_{\rm MV}\,n_{\rm GJ\star}\,r_\star^3\,L_{\gamma,1}$. This yields the quantified FP expression
\begin{equation}
\begin{split}
    L_\gamma &= C_{\rm MV} \frac{2^{4/3} q_e r_\star^{1/2}}{3^{7/3} c^{3/4} \hbar^{4/3} \pi } \eta_{\RLC}^{-2/3}\, \eta_{\alpha}^{-5/12} \, \epsilon_{\rm c}^{4/3} \, B_\star^{1/6} \, \dot{\mathcal{E}}^{5/12} \\
    &\approx 7.2 \times 10^{4} \, C_{\rm MV} \, r_6^{1/2} \, \eta_{\RLC}^{-2/3}\, \eta_{\alpha}^{-5/12} \, \epsilon_{\rm c}^{4/3} \, B_\star^{1/6} \, \dot{\mathcal{E}}^{5/12},
\end{split}
\label{eq:Lgamma CMV}
\end{equation}
where $C_{\rm MV}$ encapsulates the (effective) emitting multiplicity and volume. Equation~\ref{eq:Lgamma CMV} makes explicit the dependence on physical constants and scaling factors; the tightness of the observed FP suggests that $C_{\rm MV}$ varies only weakly across the pulsar population.

\begin{figure}
    \centering
    \includegraphics[width=1.0\linewidth]{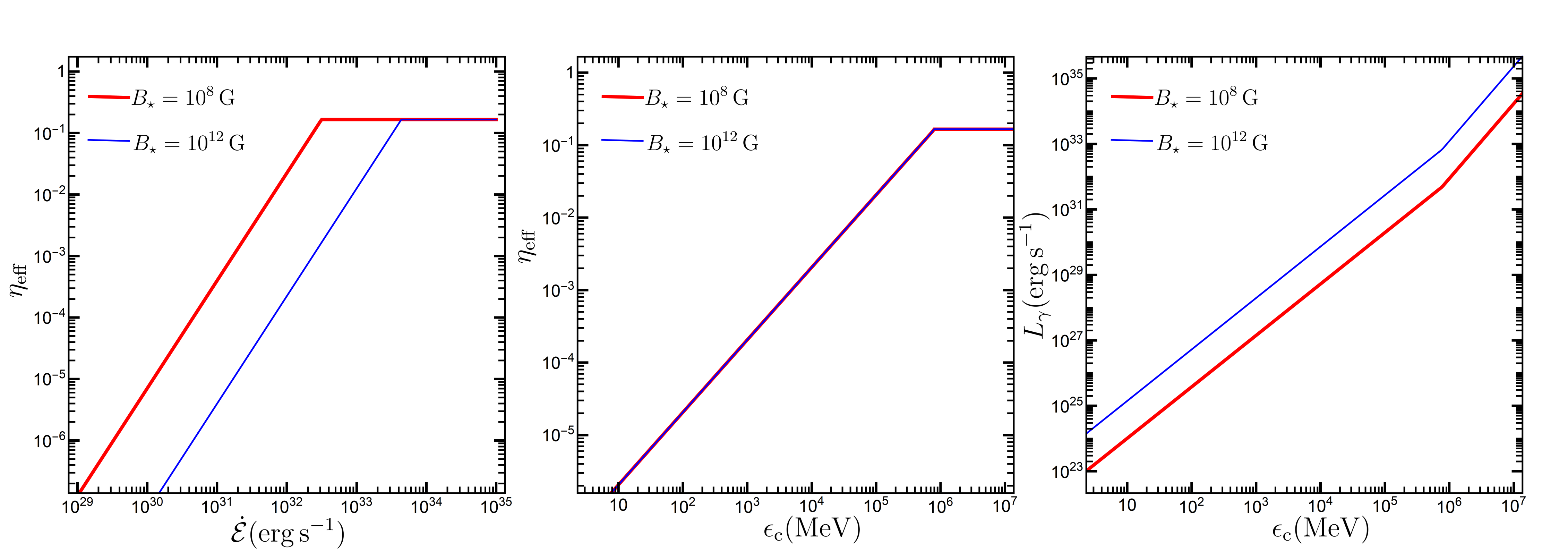}
    \caption{Behavior of gamma-ray emission across the transition from the RRL to the potential-limited regime. Each panel shows two representative curves: one for YPs ($B_\star = 10^{12}$ G, red) and one for MSPs ($B_\star = 10^{8}$ G, blue). Left-hand panel: gamma-ray efficiency $\eta_{\rm eff}$ as a function of $\dot{\mathcal E}$. Middle panel: $\eta_{\rm eff}$ versus $\epsilon_{\rm c}$ (identical for YPs and MSPs). Right-hand panel: total gamma-ray luminosity $L_\gamma$ as a function of $\epsilon_{\rm c}$.}
    \label{fig:eff-edt eff-ecut Lg-ecut}
\end{figure}

In Eq.~\ref{eq:Lgamma CMV}, the cutoff energy $\epsilon_{\rm c}$ is expressed in eV. For consistency with the observed FP relation (Eq.~\ref{eq:FP_4FGL}), we convert to MeV and obtain
\begin{equation}
    L_\gamma~({\rm erg\,s^{-1}}) \approx 7.2 \times 10^{12} \, C_{\rm MV} \, r_6^{1/2} \, \eta_{\RLC}^{-2/3}\, \eta_{\alpha}^{-5/12} \, \left(\frac{\epsilon_{\rm c}}{1 \,\, \rm MeV}\right)^{4/3}\, \left(\frac{B_\star}{1 \,\, \rm G}\right)^{1/6}\, \left(\frac{\dot{\mathcal{E}}}{1\rm  \,\,erg\,s^{-1}}\right)^{5/12}.
\label{eq:Lgamma CMV - MeV}
\end{equation}

Combining Eq.~\ref{eq:Lgamma CMV - MeV} with the RRL cutoff relation (Eq.~\ref{eq:ecut RRX}) to eliminate $\dot{\mathcal{E}}$ yields
\begin{equation}
    L_\gamma \propto \ec^{16/7}\,B_\star^{2/7}
    \label{eq: Lgamma vs ecut}
\end{equation}
for fixed scale factors and geometry. For MSPs, where $B_\star$ spans a comparatively narrow range, this implies an approximately one-parameter scaling $L_\gamma \propto \ec^{16/7}$. Interestingly, a recent phase-resolved analysis of 11 MSPs reports a tight pseudo-luminosity-cutoff-energy correlation of the form $L_{\rm p}\propto \epsilon_{\rm c}^\alpha$ with $\alpha = 2.31^{+0.22}_{-0.25}$, consistent with the curvature-radiation expectation $16/7\simeq 2.29$ \citep{2025arXiv250501708L}. Here $L_{\rm p}$ is the phase-resolved pseudo-luminosity constructed from the observed flux and distance, i.e., without an explicit beaming correction, so this agreement provides an independent test of the underlying radiative scaling up to geometric factors.

Comparing the observational FP (Eq.~\ref{eq:FP_4FGL}) with the theoretical expression in Eq.~\ref{eq:Lgamma CMV - MeV} yields $C_{\rm MV}\ \propto\ \dot{\mathcal{E}}^{\Delta a}\,\epsilon_{\rm c}^{\Delta b}\,B_\star^{\Delta c}$, where $\Delta a$, $\Delta b$, and $\Delta c$ denote the small differences between the corresponding FP exponents. Since the exponents in these two relations differ slightly within uncertainties, $C_{\rm MV}$ varies only weakly with the FP variables, i.e., it is approximately constant on the FP. This variance introduces only a mild sensitivity of $C_{\rm MV}$ to the specific values of $L_\gamma$, $\epsilon_{\rm c}$, $B_\star$, and $\dot{\mathcal{E}}$.

\begin{figure}
    \centering
    \includegraphics[width=1.0\linewidth]{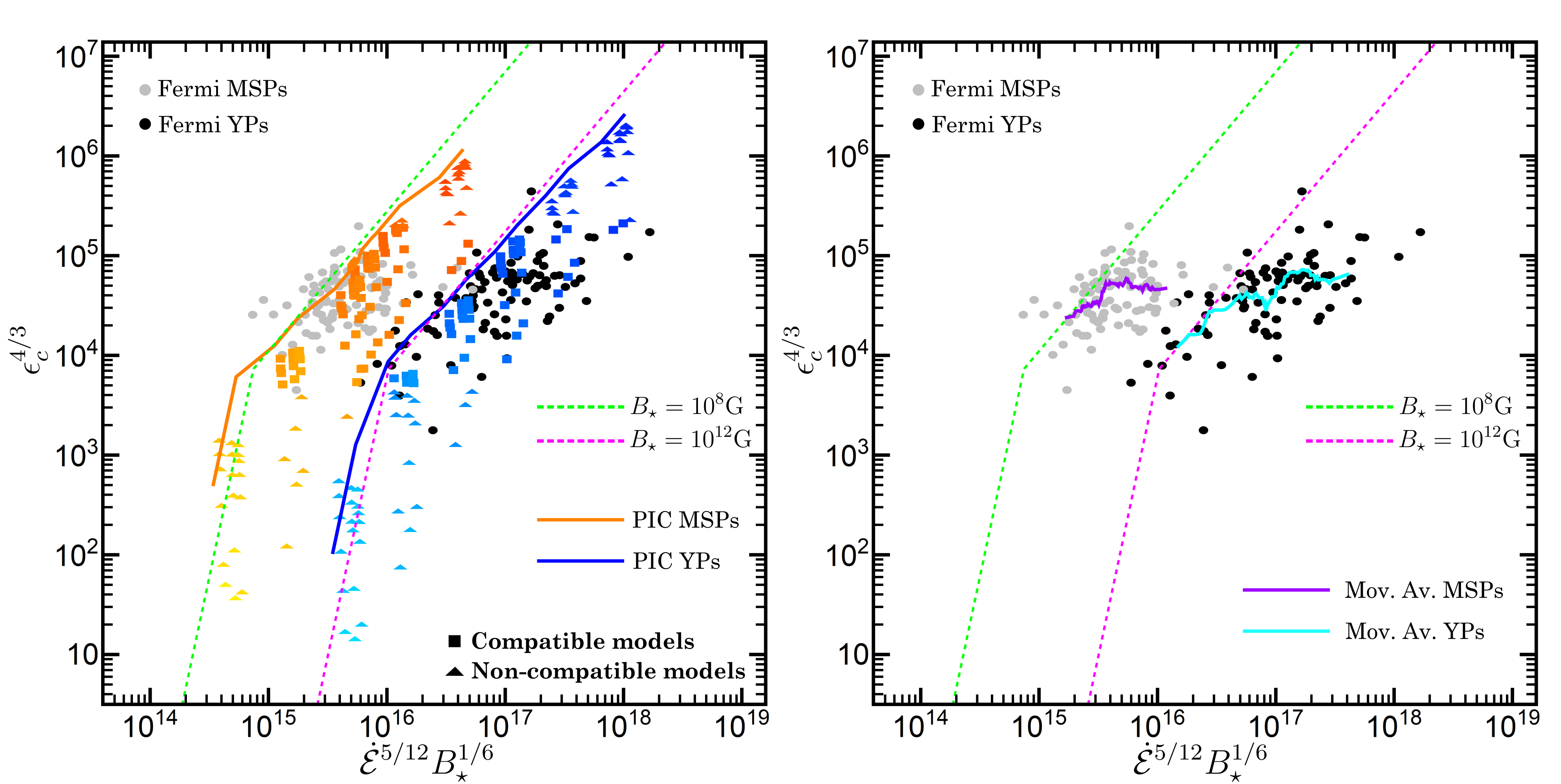}
    \caption{\textbf{Left-hand panel:} Projection of the FP showing Fermi-detected pulsars (gray: MSPs; black: YPs) together with the corresponding PIC model results (orange: MSPs; blue: YPs); for the PIC points, the color gradation encodes $L_\gamma$ (lighter shades indicate lower $L_\gamma$, sarker shades higher $L_\gamma$). Square symbols denote PIC models located within the observed FP region (compatible), while triangles indicate models without observed counterparts (non-compatible; see also \citealt{2023ApJ...954..204K}). The magenta and green dashed death lines correspond to those shown as solid lines in Fig.~\ref{fig:projectedFP Fermi pulsars and maximum ecut degeneracies} for YPs ($B_\star=10^{12}$ G) and MSPs ($B_\star=10^{8}$ G), respectively. The orange and blue solid lines show the theoretical death lines computed using the specific $B_\star$ and $\dot{\mathcal E}$ values of the PIC models. These models extend slightly below the observed range in $B_\star$ and $\dot{\mathcal E}$ while capturing the behavior of both emission regimes, indicating scaling factors consistent with those adopted in Fig.~\ref{fig:projectedFP Fermi pulsars and maximum ecut degeneracies}. \textbf{Right-hand panel:} Same FP projection including the Fermi pulsars and the death lines from the left-hand panel, now overlaid with the moving-average trends for the Fermi MSPs (purple) and YPs (cyan). The smoothed trends reveal that at lower $\dot{\mathcal E}^{5/12} B_\star^{1/6}$ values the data follow the RRL branch, while at higher values they deviate from the maximal RRL regime—consistent with enhanced screening of $E_{\rm acc}$ and an implicit dependence of $\eta_{B_{\rm LC}}$ on $\dot{\mathcal E}$.}
    \label{fig:FP projected Fermi PIC Theory - moving average trend}
\end{figure}

Using representative values, $\dot{\mathcal{E}} = 10^{35}~\rm erg\,s^{-1}$, $B_\star = 3 \times 10^{12}~\rm G$, and $\epsilon_{\rm c} = 2~\rm GeV$, and equating the right-hand sides of Eqs.~\ref{eq:FP_4FGL} and \ref{eq:Lgamma CMV - MeV}, we obtain
\begin{equation}
    C_{\rm MV} \approx 1.3\, r_6^{-1/2} \, \eta_{\RLC}^{2/3}\, \eta_{\alpha}^{5/12},
\label{eq:CMV constraint}
\end{equation}
which implies that, if the relevant scale factors (and thus $C_{\rm MV}$) are approximately constant across the FP, $M_{\rm em}$ and $f_V$ are approximately inversely proportional, $M_{\rm em}\propto f_V^{-1}$, with $C_{\rm MV}$ setting the normalization and being of order unity for our fiducial, order-unity scaling factors.

In the RRLmax regime, corresponding to Eq.~\ref{eq:ecut RRX}, the FP-based expression for the gamma-ray luminosity takes the form
\begin{equation}
\begin{split}
    L_\gamma^{\rm RRLmax} &= \frac{1}{2\pi} \, C_{\rm MV} \, \eta_{\BLC} \,  \eta_{\alpha}^{-1} \, \dot{\mathcal{E}} \\
             &\approx 0.21 \, r_6^{-1/2} \, \eta_{\BLC} \, \eta_{\RLC}^{2/3}\, \eta_{\alpha}^{-7/12} \, \dot{\mathcal{E}},
\label{eq:Lgamma rrx}
\end{split}
\end{equation}
where, in the second line, we have applied the normalization constraint from Eq.~\ref{eq:CMV constraint}.
Equation~\ref{eq:Lgamma rrx} shows that along the RRLmax efficiency boundary, the gamma-ray efficiency $\eta_{\rm eff} = L_\gamma / \dot{\mathcal{E}}$ is independent of $\dot{\mathcal{E}}$. For order-unity scaling factors, this gives $\eta_{\rm eff} \sim 20\%$, consistent with the high efficiencies observed among low-$\dot{\mathcal{E}}$ pulsars. We stress that ``RRLmax'' denotes an upper envelope within the radiation-reaction-limited regime. Pulsars may remain in the RRL regime at high $\dot{\mathcal{E}}$ while still lying below the maximal-RRL track. The clustering of many low-$\dot{\mathcal{E}}$ objects around $\eta_{\rm eff}\sim 20\%$ suggests that they approach this upper envelope, whereas higher-$\dot{\mathcal{E}}$ pulsars often fall below it, consistent with a reduction of the effective accelerating field (e.g., due to increased pair loading/screening; see Fig.~\ref{fig:FP projected Fermi PIC Theory - moving average trend} and following sections).
Furthermore, requiring $\eta_{\rm eff} \le 1$ imposes an upper bound on the combination of scaling parameters:
\begin{equation}
    \eta_{\BLC} \, \eta_{\RLC}^{2/3} \, \eta_{\alpha}^{-7/12} \lesssim 4.8 \, r_6.
\end{equation}

For the PDL regime corresponding to Eq.~\ref{eq:ecut max}, the gamma-ray luminosity along the FP takes the form
\begin{equation}
\begin{split}
    L_{\gamma}^{\rm PDL} &=\frac{q_e^5}{3\pi\, c^{39/4}\,m_e^4} r_{\star}^{-3/2} C_{\rm MV} \, \eta_{\rm pc}^{4} \, \eta_{\RLC}^{-2} \, \eta_{\alpha}^{-11/4}\, B_{\star}^{-1/2}\dot{\mathcal{E}}^{11/4}\\
    &\approx 2.8 \times 10^{-51} \, r_6^{-3/2} C_{\rm MV} \, \eta_{\rm pc}^{4} \, \eta_{\RLC}^{-2} \, \eta_{\alpha}^{-11/4}\, B_{\star}^{-1/2}\dot{\mathcal{E}}^{11/4}\\
    &\approx 3.6 \times 10^{-51} \, r_6^{-2} \, \eta_{\rm pc}^{4} \, \eta_{\RLC}^{-4/3} \, \eta_{\alpha}^{-7/3} \, B_{\star}^{-1/2}\dot{\mathcal{E}}^{11/4}\\
    &\approx 3.6 \times 10^{42} \, r_6^{-2} \, \eta_{\rm pc}^{4} \, \eta_{\RLC}^{-4/3} \, \eta_{\alpha}^{-7/3} \, B_{12}^{-1/2}\dot{\mathcal{E}}_{36}^{11/4}
    \label{eq:Lgamma maxdrop}
\end{split}
\end{equation}
which implies that, along the maximum available potential drop boundary, the gamma-ray efficiency $\eta_{\rm eff} = L_\gamma / \dot{\mathcal{E}}$ scales as $\eta_{\rm eff} \propto \dot{\mathcal{E}}^{7/4}$. Therefore, as $\dot{\mathcal{E}}$ decreases below the transition threshold $\dot{\mathcal{E}}_{\rm TR}$, the efficiency rapidly declines.

Figure~\ref{fig:eff-edt eff-ecut Lg-ecut} presents three panels illustrating the behavior of gamma-ray emission across the transition from the RRLmax regime to the PDL regime. Each panel includes two representative lines: one for YPs, i.e., $B_{\star}=10^{12}$ G (blue) and one for MSPs, i.e., $B_{\star}=10^{8}$ G (red).

The left-hand panel shows, in log-log scale, the total gamma-ray efficiency $\eta_{\rm eff}$ as a function of spin-down power $\dot{\mathcal{E}}$, capturing both the high-efficiency plateau in the RRLmax regime and the steep decline beyond the transition point. The middle panel displays $\eta_{\rm eff}$ as a function of the corresponding cutoff energy $\epsilon_{\rm c}$, again in log-log scale. In this case, the curves for YP and MSPs overlap, since the efficiency depends only on $\epsilon_{\rm c}$. The right-hand panel shows the actual gamma-ray luminosity $L_\gamma$ as a function of $\epsilon_{\rm c}$, highlighting how the luminosity evolves as emission shifts from the RRLmax regime into the PDL regime. All panels focus on the declining part of the emission behavior but include a portion of the RRLmax regime for reference. The results assume $\eta_{\RLC} = \eta_{\BLC} = r_6 = 1,\; \eta_{\rm pc} = 0.2$, and $\eta_{\alpha}=3/2$.

The sharp decline in both $\epsilon_{\rm c}$ and $\eta_{\rm eff}$ below the transition spin-down power $\dot{\mathcal{E}}_{\rm TR}$ has essential implications for the observability of pulsars occupying this region of the FP. In Section~\ref{sec:Observational consequences}, we explore this issue in greater depth, highlighting the need to develop and optimize future gamma-ray telescopes, particularly those operating in the MeV band, to access this otherwise obscured part of the pulsar population.

\section{PIC Simulations: FP Scalings, Transition, and Constraints}\label{sec:PIC Simulations}

Having established the theoretical framework for gamma-ray death lines and the two distinct emission regimes, we now turn to a direct comparison between model predictions and observations on the FP. In particular, we assess how well the behavior of PIC models of pulsars aligns with the expected phase-averaged scalings and transition structure. This comparison provides a more self-consistent (3D) numerical assessment of the theoretical framework and offers insights into the physical conditions that govern the location of pulsars on the FP, including consistent numerical values of the ``$\eta$"-scaling parameters.

In the left-hand panel of Fig.~\ref{fig:FP projected Fermi PIC Theory - moving average trend}, we project the FP onto the $\log(\dot{\mathcal{E}}^{5/12} B_\star^{1/6})$-$\log(\epsilon_{\rm c}^{4/3})$ plane, following a similar layout to Figs.~\ref{fig:projectedFP Fermi pulsars and maximum ecut}-\ref{fig:projectedFP Fermi pulsars and maximum ecut degeneracies}. Gray and black points (similar to light red and light blue points in Figs.~\ref{fig:projectedFP Fermi pulsars and maximum ecut}-\ref{fig:projectedFP Fermi pulsars and maximum ecut degeneracies}) represent the Fermi-detected MSPs and YPs, respectively. The colored points correspond to PIC simulation results from \citep[][see their Figure~14]{2023ApJ...954..204K}. Specifically, orange points denote model MSPs, and blue points denote model YPs. The model points extend well beyond the region populated by observed pulsars. Square markers denote models that lie within the observed pulsar region, while triangular markers represent models located in currently unobserved areas of the FP. In \citet{2023ApJ...954..204K}, the former were referred to as ``compatible" and the latter as ``non-compatible" models.

In the same figure, we overlay analytic curves representing characteristic scaling-factor combinations. The green and magenta lines correspond to $B_\star = 10^8$ G (MSP branch) and $B_\star = 10^{12}$ G (YP branch), respectively, using $\eta_{\RLC}=1$, $\eta_{\BLC}=1$, $\eta_{\rm pc}=0.2$, $\eta_{\alpha}=3/2$, and $r_6=1$. These curves provide a coarse theoretical description based on representative magnetic-field values and reproduce the general envelope of the observed and PIC model behavior. We note that multiple combinations of $(\eta_{\RLC},\eta_{\BLC},\eta_{\rm pc})$ can reproduce the same envelop (see Fig.~\ref{fig:projectedFP Fermi pulsars and maximum ecut degeneracies}).

To further demonstrate the agreement between the PIC results and the theoretical framework, we overplot in orange and blue the theoretical curves derived directly from the specific model inputs, namely, the six $B_\star$ values and the corresponding simulation $\dot{\mathcal{E}}$ values for MSPs and for YPs in \citet{2023ApJ...954..204K}, together with their corresponding transition points $\dot{\mathcal{E}}_{\rm TR}$. These lines do not fit the simulation data; they are obtained by evaluating the analytic relations at the simulation parameters. The close correspondence between the PIC model points and these theoretical curves reinforces the consistency of the framework and its ability to reproduce the overall scaling behavior of both pulsar populations.

The envelope formed by the highest model $\epsilon_{\rm c}$ values delineates the transition between the RRLmax and PDL regimes and captures the characteristic behavior within each. In the RRLmax regime, $\epsilon_{\rm c}\propto \dot{\mathcal{E}}^{7/16}$; in the PDL regime, $\epsilon_{\rm c}\propto \dot{\mathcal{E}}^{7/4}$. The break between these slopes identifies the transition spin-down power, $\dot{\mathcal{E}}_{\rm TR}$. Models lying closest to this upper envelope correspond to the lowest particle injection rates in the separatrix layer, which favor the highest achievable $\epsilon_{\rm c}$ values.

In the PIC simulations, as in real pulsars, the emission arises from particle distributions that sample a range of local conditions, effectively corresponding to different scaling factors. Nevertheless, the global behavior is well captured by a set of \emph{effective} scale factors. Because the upper envelope reproduces both the regime-specific slopes and the transition location, it encodes these effective combinations and, despite underlying parameter degeneracies, delineates robust constraints on their allowed values.

While the upper envelope of the observed pulsars’ $\epsilon_{\rm c}$ values appears to follow the theoretical RRLmax trend reasonably well, despite some scatter in the data, we examine this more closely in the right-hand panel of Fig.~\ref{fig:FP projected Fermi PIC Theory - moving average trend}. For clarity, we replot the projected FP, showing the Fermi pulsars as gray circles (MSPs) and black circles (YPs), along with the theoretical guidelines (for $B_{\star}=10^8$ G and $B_{\star}=10^{12}$ G) from the left-hand panel. 

To highlight overall trends, we overlay moving-average lines for both populations: purple for MSPs and cyan for YPs. Here, the ``moving-average'' trend is computed separately for MSPs and YPs by sorting sources by the projected-FP abscissa and then applying a sliding window of $N=15$ sources. In each window we compute both the moving mean and the moving median of the corresponding ordinate (in log-space), and we plot their average as our smoothed trend. We have verified that using different window sizes does not change the qualitative behavior, aside from the expected increase in noise for smaller $N$. At low $\dot{\mathcal{E}}$, these moving averages align well with the theoretical RRLmax lines. However, in both populations, we observe an apparent deviation from the RRLmax trend occurring at similar values of $\epsilon_{\rm c}$. Beyond this point, at higher $\dot{\mathcal{E}}$, the observed $\epsilon_{\rm c}$ values flatten out and fall below the RRLmax envelope. Within our framework, this behavior directly implies that the effective accelerating field in the ECS does not remain at its maximal value, i.e., $\eta_{B_{\rm LC}}$ must decrease below $\eta_{B_{\rm LC}}^{\max}$, consistent with the onset of enhanced pair production that screens $E_{\rm acc}$ in the emitting/accelerating region. Quantitatively, enforcing an approximately stabilized $\epsilon_{\rm c}$ above the deviation point via Eq.~\ref{eq:ecut RRX} requires $\eta_{B_{\rm LC}}\propto \dot{\mathcal{E}}^{-7/12}$, very close to the behavior inferred in earlier ECS-based studies \citep[e.g.,][]{2017ApJ...842...80K,2022ApJ...934...65K}. We explore and discuss the origin and implications of this deviation in Section~\ref{sec:Discussion Conclusions}.

\begin{figure}
    \centering
    \includegraphics[width=1.0\linewidth]{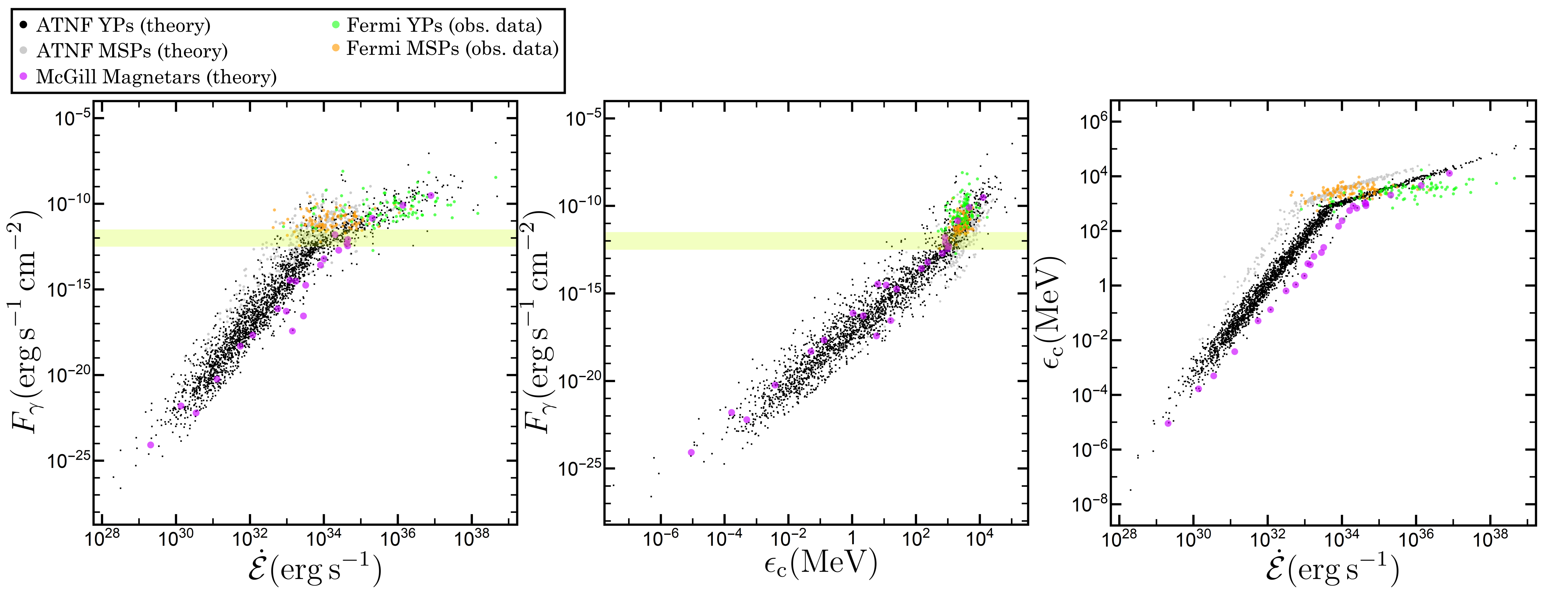}
    \caption{\textbf{Left-hand panel:} Predicted gamma-ray flux at Earth, $F_{\gamma}$, versus spin-down power, $\dot{\mathcal{E}}$, for pulsars from the ATNF catalog (gray: MSPs; black: YPs), Fermi-detected pulsars from \citet{2022ApJ...934...65K} (orange: MSPs; green: YPs), and magnetars from the McGill catalog (magenta). For Fermi pulsars, measured $F_{\gamma}$ values are plotted, while for ATNF pulsars and magnetars, $F_{\gamma}$ is computed aasuming $L_{\gamma}$ values from Eqs.~\ref{eq:Lgamma rrx} and \ref{eq:Lgamma maxdrop}, depending on whether $\dot{\mathcal{E}}$ lies above or below $\dot{\mathcal{E}}_{\rm TR}$ (Eq.~\ref{eq:transition spin-down B}), assuming a beaming factor $f_b = 1$ and the scaling factors corresponding to the death lines shown in Figs.~\ref{fig:projectedFP Fermi pulsars and maximum ecut degeneracies} and \ref{fig:FP projected Fermi PIC Theory - moving average trend}. The yellow band marks the Fermi-LAT sensitivity range. \textbf{Middle panel:} $F_{\gamma}$ versus $\epsilon_{\rm c}$. For Fermi pulsars, the reported $\epsilon_{\rm c1}$ values \citep{2022ApJ...934...65K} are shown; for ATNF pulsars and magnetars, $\epsilon_{\rm c}$ values are computed from the Eqs.~\ref{eq:ecut RRX} and \ref{eq:ecut max} depending again on the corresponding regime. \textbf{Right-hand panel:} $\epsilon_{\rm c}$ versus $\dot{\mathcal{E}}$ for the same sources, combining information from the two other panels. These diagrams illustrate the sensitivity limits and energy ranges relevant for detecting presently undetected pulsar classes. Sources predicted above the  {Fermi}-LAT sensitivity but not detected are likely overestimated due to the assumption of the maximal RRL regime at high $\dot{\mathcal{E}}$ (see Figs.~\ref{fig:FP projected Fermi PIC Theory - moving average trend}, \ref{fig:Fg_edot_ec=1GeV}, and text), or have unfavorable geometries that prevent their gamma-ray beams from reaching Earth.}
    \label{fig:edot-Fg_ecut-Fg_edot-ecut}
\end{figure}

\section{Observational consequences}\label{sec:Observational consequences}

Our previous analysis (e.g., see Fig.~\ref{fig:FP projected Fermi PIC Theory - moving average trend}) indicates that the spectral $\ec$ values closely follow the maximum available ones below some spin-down power, i.e., the point where the moving average values start deviating from the RRLmax lines. 

To assess how pulsar detectability by current and future gamma-ray instruments is influenced by the emission behaviors described above, we consider an \textit{upper-limit} scenario in which all pulsars emit at their maximum achievable $\epsilon_{\rm c}$. This assumption is particularly justified at lower $\dot{\mathcal{E}}$ values, below which the observed moving-average $\epsilon_{\rm c}$ matches the maximum RRLmax regime. Specifically, we considered all pulsars ($\sim 2750$) from the ATNF catalog with recorded values of spin-down power ($\dot{\mathcal{E}}$), surface magnetic field ($B_{\star}$), period ($P$), and distance ($d$)\footnote{These are predominantly radio-loud pulsars, as reliable distance estimates are generally unavailable for radio-quiet systems.}. For each pulsar, we computed the corresponding maximum cutoff energy ($\epsilon_{\rm c}$) using either Eq.~\ref{eq:ecut RRX} (RRLmax regime) or Eq.~\ref{eq:ecut max} (PDL regime), based on its $\dot{\mathcal{E}}$, assuming scaling factors of $\eta_{\RLC}=1$, $\eta_{\BLC}=1$, and $\eta_{\rm pc}=0.2$. With these calculated $\epsilon_{\rm c}$ values and adopted $\eta$ factors, we then derived the corresponding gamma-ray luminosity ($L_{\gamma}$) from Eqs.~\ref{eq:Lgamma rrx} and \ref{eq:Lgamma maxdrop}. Finally, using the known distances and assuming an Earth-directed beaming factor of $f_b=1$ \citep{2010ApJ...714..810R}, we calculated the corresponding gamma-ray energy flux, $F_{\gamma}= L_{\gamma}/(4 \pi d^2)$, at Earth for each pulsar.

\begin{figure}
    \centering
    \includegraphics[width=0.5\linewidth]{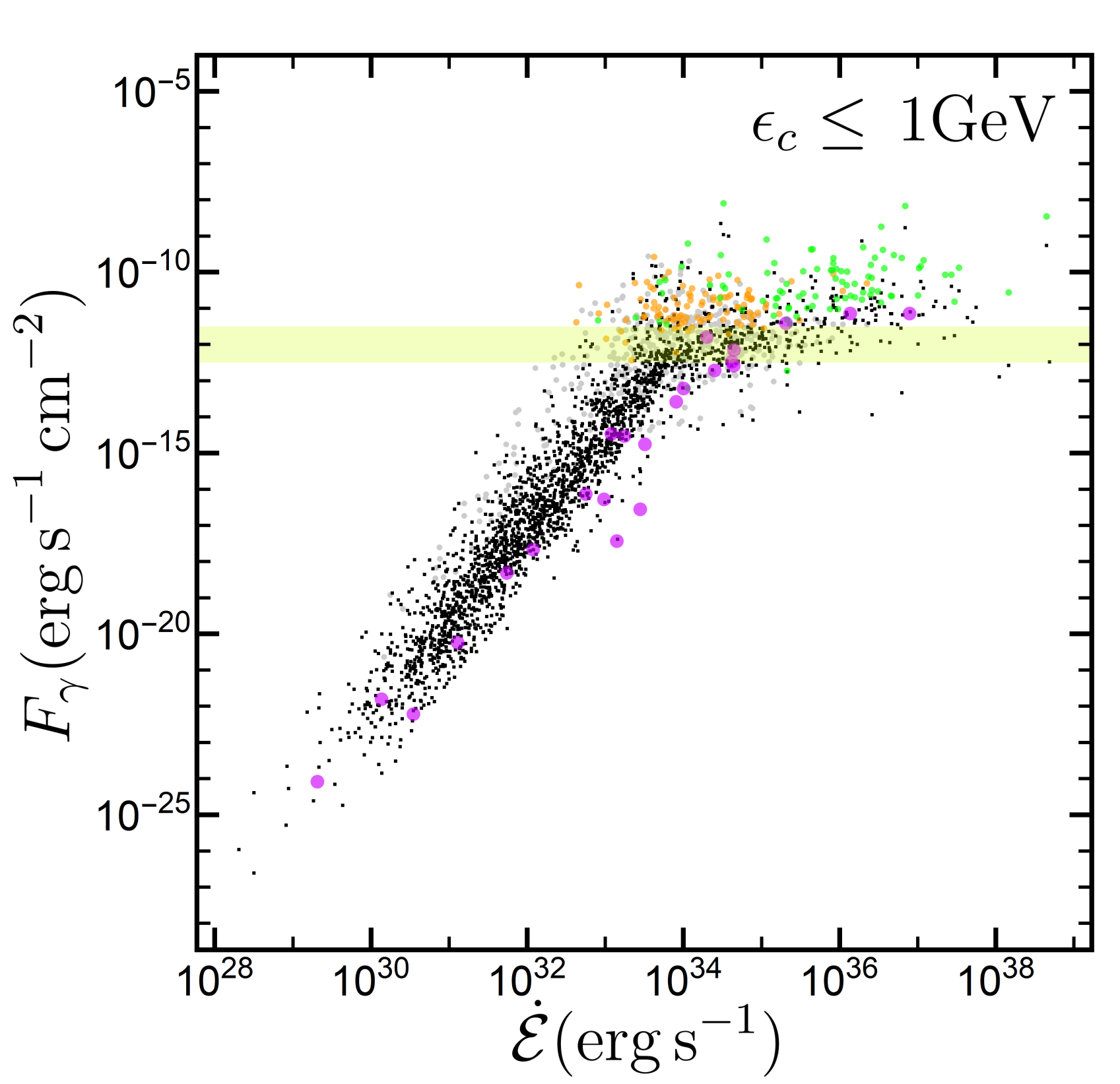}
    \caption{Same as the left-hand panel of Fig.~\ref{fig:edot-Fg_ecut-Fg_edot-ecut}, but imposing an upper limit of $1$~GeV on $\epsilon_{\rm c}$ to mimic the observed flattening of the cutoff energy, plausibly associated with the onset of additional pair regulation. Under this constraint, the predicted $F_\gamma$ values for the highest-$\dot{\mathcal{E}}$ objects (gray, black, and magenta) are systematically reduced relative to Fig.~\ref{fig:edot-Fg_ecut-Fg_edot-ecut}, pushing them closer to (or below) detectability thresholds and making their detection more challenging.}
    \label{fig:Fg_edot_ec=1GeV}
\end{figure}

In addition to the ATNF catalog, we also included magnetars from the McGill magnetar catalog to ensure a more complete representation of the high-magnetic-field neutron star population. While some magnetars are already listed in the ATNF catalog, the McGill catalog provides an updated and dedicated compilation, albeit with a relatively small sample size (about 30 objects). Consequently, the overlap between the two catalogs is minimal and does not affect the overall statistics. Any duplicate entries were retained without modification, since they are inconsequential to the analysis.

In the left-hand panel of Fig.~\ref{fig:edot-Fg_ecut-Fg_edot-ecut}, we present a $\log$-$\log$ plot of $F_{\gamma}$ versus $\dot{\mathcal{E}}$. The gray and black points represent MSPs ($B_{\star}<10^{10}$~G) and YPs ($B_{\star}>10^{10}$~G), respectively, from the ATNF catalog.\footnote{In this study, we classify pulsars based on their magnetic field strength, in contrast to other works that adopt a period threshold of 30~ms. We find that using $B_{\star}$ provides a more effective separation of their emission properties.} The magenta points denote magnetars from the McGill catalog. The orange and green points correspond to Fermi MSPs and YPs, respectively, taken from \citep{2022ApJ...934...65K}. We selected the Fermi pulsar sample from \cite{2022ApJ...934...65K}, which is based on 4FGL catalog data \citep{2022ApJS..260...53A}, rather than directly using \citetalias{Smith23}, because that study derived $\epsilon_{c1}$ values, the cutoff energy parameter corresponding to a pure exponential cutoff (plus a power-law) spectral form. This parameter provides an optimal probe of the intrinsic maximum cutoff energy associated with emission from the core of the dissipative region (the ECS). It is worth noting that the underlying data in \citetalias{Smith23} also originate from the 4FGL catalog. For the Fermi pulsars, we adopt the measured $\epsilon_{c1}$ and $L_{\gamma}$ values, rather than the maximal emission values (corresponding to the RRLmax or PDL regimes) used for the sources in the ATNF and McGill catalogs. 
The transparent yellow-shaded region marks the effective sensitivity range of Fermi-LAT, approximately $10^{-11.5}$ to $10^{-12.5}~\mathrm{erg~cm^{-2}~s^{-1}}$ (which depends on sky coordinates).

The entire ATNF group clearly shows two distinct regimes: RRLmax and PDL. The Fermi pulsars lie near the RRLmax regime (and start deviating from it for larger $\ed$), and their flux reaches the LAT sensitivity threshold just before the knee, corresponding to the transition between the two regimes. This is not a fine-tuned coincidence. In our framework the knee marks an intrinsic transition after which both the cutoff energy and the radiative efficiency drop rapidly on the PDL branch, so for typical distances the predicted GeV flux falls below the LAT sensitivity soon after the transition, consistent with the regime change suggested by the PIC trends. Consequently, the lowest-$\dot{\mathcal{E}}$ LAT detections primarily provide an upper limit on how high $\dot{\mathcal{E}}_{\rm TR}$ can be, while the detailed confirmation requires selection effects and population synthesis. The magnetars similarly populate both regimes; however, only three appear to lie clearly above the LAT sensitivity threshold.

In the middle and right-hand panels of Fig.~\ref{fig:edot-Fg_ecut-Fg_edot-ecut}, we show $\log$-$\log$ plots of $F_\gamma$ versus $\epsilon_{\rm c}$ and $\epsilon_{\rm c}$ versus $\dot{\mathcal{E}}$, respectively, using the same color coding as in the left-hand panel. Both panels reveal the two emission regimes, with their distinction most evident in the right-hand panel. In the middle panel, the LAT sensitivity threshold lies near the transition knee between the regimes. At high fluxes, within the RRLmax regime, several ATNF catalog pulsars and a few magnetars exhibit $\epsilon_{\rm c}$ values at or beyond those detected by Fermi-LAT; this divergence is even more apparent in the right-hand panel, where the ATNF and magnetar $\epsilon_{\rm c}$ values depart from the Fermi-LAT trend at high $\dot{\mathcal{E}}$. This behavior arises because the ATNF and magnetar values reflect theoretical maxima. Consistent with Fig.~\ref{fig:FP projected Fermi PIC Theory - moving average trend}, the ATNF values closely follow the RRLmax prediction at low $\dot{\mathcal{E}}$, but deviate at higher $\dot{\mathcal{E}}$, exhibiting lower $\epsilon_{\rm c}$ values than those expected from the RRLmax scaling.

\begin{figure}
    \centering
    \includegraphics[width=1.0\linewidth]{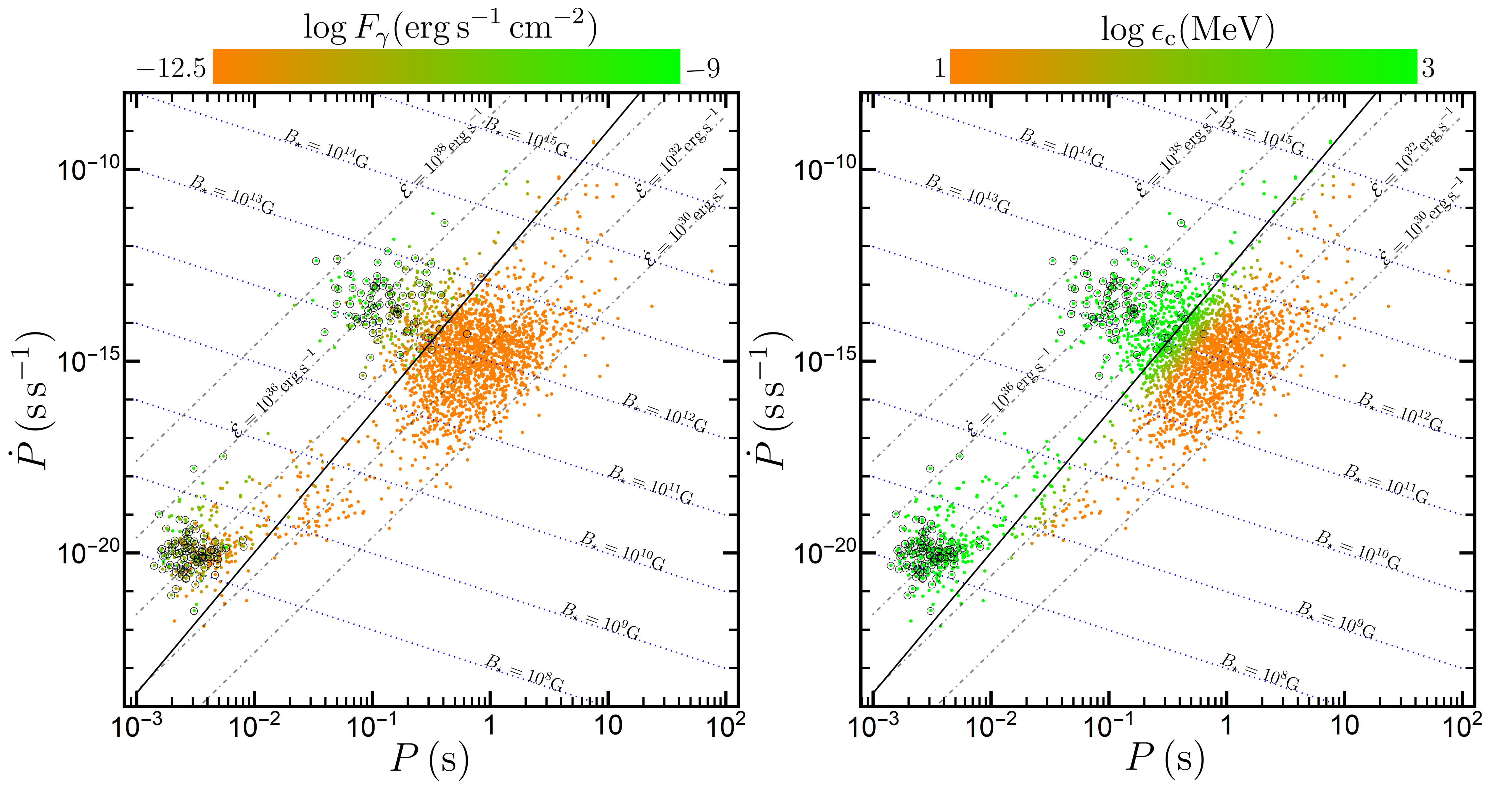}
    \caption{The $P$-$\dot{P}$ diagram for all pulsars in our combined ATNF and McGill sample. Fermi-detected pulsars are marked with open circles. Blue dotted and black dot-dashed lines indicate constant $B_{\star}$ and $\dot{\mathcal{E}}$ values, respectively, while the thick black curve marks the ``visibility line'' where $\dot{\mathcal{E}}=\dot{\mathcal{E}}_{\rm TR}$ (Eq.~\ref{eq:PPdot_death_line}), for $\eta_{R_{\rm LC}}=1$, $\eta_{B_{\rm LC}}=1$, $\eta_{R_{\rm pc}}=0.2$, and $\eta_{\alpha}=1$. The color scale shows $\log F_{\gamma}$ in the left-hand panel and $\log \epsilon_{\rm c}$ in the right-hand panel. For Fermi pulsars, the observed $F_{\gamma}$ and $\epsilon_{\rm c}$ values are used \citep{2020ApJS..247...33A,2022ApJS..260...53A,2022ApJ...934...65K}; for the rest, these quantities are computed from Eqs.~\ref{eq:Lgamma rrx}, \ref{eq:Lgamma maxdrop}, \ref{eq:ecut RRX}, and \ref{eq:ecut max}. In the left panel, both greenish and orangish points appear above the visibility line, green where sources are bright enough to be detectable, and orange where distance limits detection, while only orange points remain below it, reflecting the rapid decrease in $F_{\gamma}$ and efficiency. The sole outlier, PSR~J2208+4056, lies slightly below the line yet is detected (see text). In the right panel, greenish points mark sources with $\epsilon_{\rm c}\gtrsim100$~MeV, within the Fermi-LAT band. The visibility line encompasses most MSPs, bisects the YP population, and leaves magnetars below it—consistent with their non-detection in gamma rays. Where the visibility line crosses the YP locus roughly midway, the lowest-detected YPs closely track the threshold, providing empirical validation of the boundary.}
    \label{fig:P Pdot visibility line}
\end{figure}

An important implication from these figures, and consistent with the theoretical analysis presented earlier, is that there is no intrinsic reason why MSPs and YPs should cease emitting rotation-powered high-energy radiation at lower $\dot{\mathcal{E}}$ provided plasma is able to be supplied from polar cap pair cascades, i.e., the pulsar remains above the radio death line. However, the sharp decline (below $\ed_{\rm TR}$) in both $\epsilon_{\rm c}$ and the corresponding lower gamma-ray efficiency, combined with the intrinsically lower $\dot{\mathcal{E}}$ values, makes the detection of high-energy emission from such pulsars increasingly difficult at low spin-down powers.

The middle column of Fig.~\ref{fig:edot-Fg_ecut-Fg_edot-ecut} further illustrates the sensitivity requirements necessary to detect pulsars across photon energies, i.e., as a function of $\epsilon_{\rm c}$, and $\dot{\mathcal{E}}$. We emphasize, however, that these sensitivity estimates are primarily applicable to the higher-$\dot{\mathcal{E}}$ objects among the currently undetected population. It remains uncertain how far down in $\dot{\mathcal{E}}$ the same emission conditions and assumptions can be extrapolated.

Figure~\ref{fig:edot-Fg_ecut-Fg_edot-ecut} shows that several MSPs, YPs, and possibly a few magnetars fall within the observable flux region yet remain undetected in gamma rays. Several factors could account for this. First, geometric beaming effects may prevent some radio pulsars from being observable in gamma rays, regardless of their intrinsic luminosity. While lower $\dot{\mathcal{E}}$ pulsars tend to be fainter on average, a misaligned or narrowly directed gamma-ray beam can make even intrinsically bright sources undetectable to an Earth-based observer. Second, as discussed earlier, above the characteristic transition spin-down power $\dot{\mathcal{E}}_{\rm TR}$ for each population, pulsars may no longer operate at their maximum available $\epsilon_{\rm c}$. Although they remain in the RRL regime, increased pair-production efficiency reduces the available accelerating electric field, leading to lower $\epsilon_{\rm c}$ values and, consequently, lower $L_\gamma$. While these pulsars still follow the FP relation, their $L_\gamma$ reflects the reduced $\epsilon_{\rm c}$, resulting in both lower $\epsilon_{\rm c}$ and reduced gamma-ray efficiency, making detection more challenging. To illustrate this effect, Fig.~\ref{fig:Fg_edot_ec=1GeV}, analogous to the left-hand panel of Fig.~\ref{fig:edot-Fg_ecut-Fg_edot-ecut}, shows the expected gamma-ray fluxes for ATNF and magnetar sources with $\dot{\mathcal{E}} > \dot{\mathcal{E}}_{\rm TR}$, assuming a fixed $\epsilon_{\rm c} = 1~\mathrm{GeV}$ for all. In this scenario, most sources shift closer to, or below, the detection threshold (yellow zone).

\subsection{The Gamma-Ray Visibility Threshold on the $P$-$\dot{P}$ Diagram}

Our analysis so far indicates that while the transition spin-down power, $\dot{\mathcal{E}}_{\rm TR}$, is not a strict cutoff below which gamma-ray emission ceases, pulsar detectability rapidly diminishes below this point. Thus, $\dot{\mathcal{E}}_{\rm TR}$ effectively acts as an inflection point in the gamma-ray emission evolution of pulsars, marking the regime where luminosities and cutoff energies sharply decline. To clearly connect this theoretical threshold with observational parameters commonly used to classify pulsars, we explicitly derive the corresponding gamma-ray pulsar death line on the standard $P$-$\dot{P}$ diagram below.

To further illustrate the implications of our findings within a familiar observational framework, we explicitly express the gamma-ray pulsar death line in terms of the standard pulsar observables $P$ and $\dot{P}$. Starting from the definition of spin-down power, $\dot{\mathcal{E}} = 4\pi^2 I_{\rm ns}\dot{P}/P^3$
and using our previously derived transition spin-down power $\dot{\mathcal{E}}_{\rm TR}$ (Eq.~\ref{eq:transition spin-down P}), we equate these expressions at the death line condition:
\begin{equation}
    4\pi^2 I_{\rm ns}\frac{\dot{P}_{\rm vis}}{P^3} 
    = 1.2\times10^{32}~P^{2/3}\eta_{\RLC}^{4/3}\eta_{\BLC}^{2/3}\eta_{\rm pc}^{-8/3}\eta_{\alpha}.
\end{equation}
Solving explicitly for $\dot{P}_{\rm vis}$ and adopting the standard neutron star moment of inertia ($I_{\rm ns}=10^{45}\,\mathrm{g\,cm^2}$), we obtain the gamma-ray pulsar \emph{visibility} line on the $P$-$\dot{P}$ plane:
\begin{equation}
    \dot{P}_{\rm vis}\approx 3\times10^{-15}\,\eta_{\RLC}^{4/3}\eta_{\BLC}^{2/3}\eta_{\rm pc}^{-8/3}\,\eta_{\alpha}\,P^{11/3}.
    \label{eq:PPdot_death_line}
\end{equation}
This relation provides a clear observational boundary that marks the transition between practically detectable gamma-ray pulsars and those whose emission is effectively suppressed by insufficient acceleration potential or strong radiation-reaction effects. Pulsars located below or near this line in the $P$-$\dot{P}$ diagram exhibit sharply reduced gamma-ray luminosities and cutoff energies, challenging detection with current gamma-ray observatories.

In Fig.~\ref{fig:P Pdot visibility line}, we present the $P$-$\dot{P}$ diagram (in log-log scale) for all pulsars in our combined ATNF and McGill sample. Both panels display the same set of objects but differ in the quantity represented by the color scale. In the left-hand panel, colors indicate the predicted gamma-ray energy flux at Earth, $F_{\gamma}$ (corresponding to the maximum $L_{\gamma}$ values from Eqs.~\ref{eq:Lgamma rrx} and \ref{eq:Lgamma maxdrop}). In contrast, in the right-hand panel, they represent the predicted spectral cutoff energy, $\epsilon_{\rm c}$ (see Eqs.~\ref{eq:ecut RRX} and \ref{eq:ecut max}). The color scale ranges from orange to green, corresponding to $\log F_{\gamma} = -12.5$ to $-9$ in the left-hand panel and $\log \epsilon_{\rm c} = 1$ to $3$ ($\ec$ in MeV) in the right-hand panel, as indicated in the figure. Pulsars detected by Fermi-LAT are shown with open circles, and the solid black curve marks the gamma-ray visibility threshold derived from Eq.~\ref{eq:PPdot_death_line}, corresponding to $\dot{\mathcal{E}} = \dot{\mathcal{E}}_{\mathrm{TR}}$.

In the right-hand panel, greenish points mark sources with $\epsilon_{\rm c}\gtrsim 100~\mathrm{MeV}$, i.e., within the Fermi-LAT band. All but one detected pulsars lie on the observable side of the visibility curve. Greenish points that appear on the non-observable side indicate sources that could, in principle, produce $> 100$~MeV photons but whose predicted $F_{\gamma}$ falls below the Fermi-LAT sensitivity, consistent with their positions and colors in the left-hand panel.

In the left-hand panel, the region above the visibility line exhibits both green and orange points: green where sources are bright enough to be detectable and orange where large distances render otherwise observable systems undetectable. Below the visibility line, we see only orange points, reflecting the rapid drop of $F_{\gamma}$ (and, concomitantly, $\epsilon_{\rm c}$ and efficiency) and the consequent lack of detections. There is one notable exception: PSR~J2208+4056, which lies slightly below the line yet is detected. Its radio interpulse geometry implies an equatorial line of sight and hence a beaming factor $f_b<1$ \citep[][]{2019ApJ...871...78S} (so that $F_\gamma = L_\gamma/(4\pi f_b d^2)$), which boosts the Earthward flux relative to the $f_b = 1$ assumption used in Fig.~\ref{fig:P Pdot visibility line}. This naturally explains why J2208+4056 lies slightly below the visibility locus yet is detected by Fermi-LAT.

Where the visibility line crosses the YP locus roughly midway, the lowest-detected YPs track the threshold closely (again, with the sole outlier PSR~J2208+4056 just below it), providing empirical validation of the boundary. The threshold places the vast majority of MSPs on the observable side, cuts through the YP population approximately in the middle, and leaves most magnetars on the non-observable side, consistent with their current non-detection in gamma rays.

It is important to emphasize that a pulsar’s position relative to the visibility threshold does not by itself guarantee detectability. Beaming geometry and distance are critical: green points above the line that remain undetected are plausibly due to unfavorable beaming, whereas orange points above the line are generally too distant. Conversely, below the line, there are no green points in the left-hand panel precisely because the flux drops quickly; detection would require exceptionally small distances, which are rare. In this sense, Eq.~\eqref{eq:PPdot_death_line} defines a \emph{necessary} condition for practical detectability, not a sufficient one. A full population-synthesis treatment that samples the underlying $P$–$\dot P$–distance–beaming distributions will naturally place many more MSPs and YPs above the visibility line than are currently detected, and reconciling that abundance with the modest LAT sample requires incorporating both the empirical flattening/saturation of $\epsilon_{\rm c}$ at a few GeV and realistic selection effects, as explored in a companion study \citep{gammaPTA}.

Interestingly, we also note green-colored points (i.e., high predicted flux) above the visibility threshold that remain undetected. This subset, most notably a cluster of MSPs at high $B_\star$ and high $\dot{\mathcal{E}}$, may reflect practical limitations (e.g., orbital-period wander and eclipses in compact binaries that disrupt radio timing solutions, or a bright diffuse gamma-ray background) rather than intrinsic faintness. Improved radio timing solutions and extended ephemerides, enabling phase-coherent gamma-ray searches, could materially affect their detectability. We therefore flag these objects as priority candidates for timing follow-up and targeted Fermi-LAT pulsation searches.

As discussed above, Fig.~\ref{fig:P Pdot visibility line} was built under an upper-limit (optimistic) assumption that each source radiates at its maximum available $\epsilon_{\rm c}$ and corresponding $L_\gamma$. In practice, enhanced pair production at high $B_\star$ and high $\dot{\mathcal{E}}$ reduces the effective accelerating electric field in the ECS, lowering $\epsilon_{\rm c}$ (still within the RRL regime) and thus $L_\gamma$ and $F_\gamma$. This correction is  {most consequential for magnetars}: the highest-$\dot{\mathcal{E}}$ magnetars already sit close to the visibility threshold, so any reduction in $\epsilon_{\rm c}$ or $L_\gamma$ readily pushes them below detectability. By contrast, high-$\dot{\mathcal{E}}$ YPs and MSPs generally lie farther above the threshold; for them, geometry (beaming) and distance typically dominate the detection outcome, with the same pair-induced suppression shifting them down but not necessarily past the threshold. When one moves from the catalog-based exercise here to a synthetic Galactic population, this saturation and suppression of $\epsilon_{\rm c}$ are essential ingredients for avoiding an overabundance of predicted LAT detections at high $\dot{\mathcal{E}}$.

The significant reduction in pulsar gamma-ray luminosity and spectral cutoff energy at lower spin-down powers emphasizes the need for sensitive observational facilities in the MeV energy range. At photon energies below $\sim 1$--10~MeV, however, synchrotron components from pairs and primaries are expected to dominate the flux over the curvature component that underlies our FP-based visibility analysis \citep[e.g.,][]{2015ApJ...811...63H,2018ApJ...869L..18H,2021ApJ...923..194H}. For most LAT pulsars, these synchrotron components carry a smaller fraction of the total high-energy power than the GeV curvature peak, with the Crab being a notable exception. In this low-energy regime, synchrotron emission from primary and secondary pairs, whether produced inside the light cylinder or in the ECS reconnection layers, will generally set the observable spectrum. Thus, the visibility line derived here should be interpreted as a condition on where the high-energy ECS accelerator remains efficient and can sustain bright MeV--GeV core emission, rather than as a detailed model of the pair-synchrotron output at the lowest MeV energies. 

Future telescopes, such as the planned All-sky Medium Energy Gamma-ray Observatory eXplorer (AMEGO-X) \citep{2022JATIS...8d4003C} and the Gamma-Ray and AntiMatter Survey (GRAMS) Compton/pair telescope concept \citep{2020APh...114..107A}, are specifically designed to bridge the current observational gap between keV X-ray and GeV gamma-ray instruments. A smaller, nearer-term forerunner is the Compton Spectrometer and Imager (COSI) mission \citep{2024icrc.confE.745T}. With improved sensitivity and a broader effective energy range, AMEGO-X and similar next-generation missions that utilize advanced detection technologies, such as Time Projection Chambers (TPCs) \citep{2025arXiv250214841S}, will be crucial both for detecting the predicted MeV-dominated, GeV-faint pulsar, magnetar, and other compact objects below the LAT sensitivity threshold and for mapping the pair-synchrotron components in already LAT-detected pulsars. Such MeV-focused missions could directly test theoretical predictions regarding emission regimes, death lines, and transitions described above, significantly enhancing our understanding of pulsar emission physics and population characteristics.

\section{The TeV Photon regime}\label{sec:The TeV regime}

In this section, we examine how the theoretical framework developed above, particularly as realized through our PIC simulations \citep{2023ApJ...954..204K}, can account not only for the GeV gamma-ray emission observed by Fermi-LAT, but also for the recently discovered pulsed TeV emission from the Vela pulsar \citep{2022tsra.confE..33D,2023NatAs...7.1341H}. To this end, we adopt a simplified model for the target photon population, following proposals in the literature, and apply it to a set of PIC models characterized by varying degrees of force-free-ness (FF-ness). These variations yield different particle energy distributions and corresponding curvature radiation spectra. For each case, we compute the associated IC scattering spectrum and directly compare both the curvature and IC components to the observed Fermi-LAT and H.E.S.S. II spectra of Vela.

While this analysis is intended as a demonstration rather than a detailed emission model, it highlights that the framework developed in this work is not only capable of reproducing the observed GeV and TeV spectra but also appears fundamentally necessary for doing so in a self-consistent manner. The connection between curvature radiation in the ECS and IC upscattering of ambient photons emerges naturally from the structure and energetics of the PIC simulations, suggesting that a unified treatment of pulsar magnetospheric physics is essential for explaining the full extent of high-energy emission.

\begin{figure}
    \centering\includegraphics[width=0.5\linewidth]{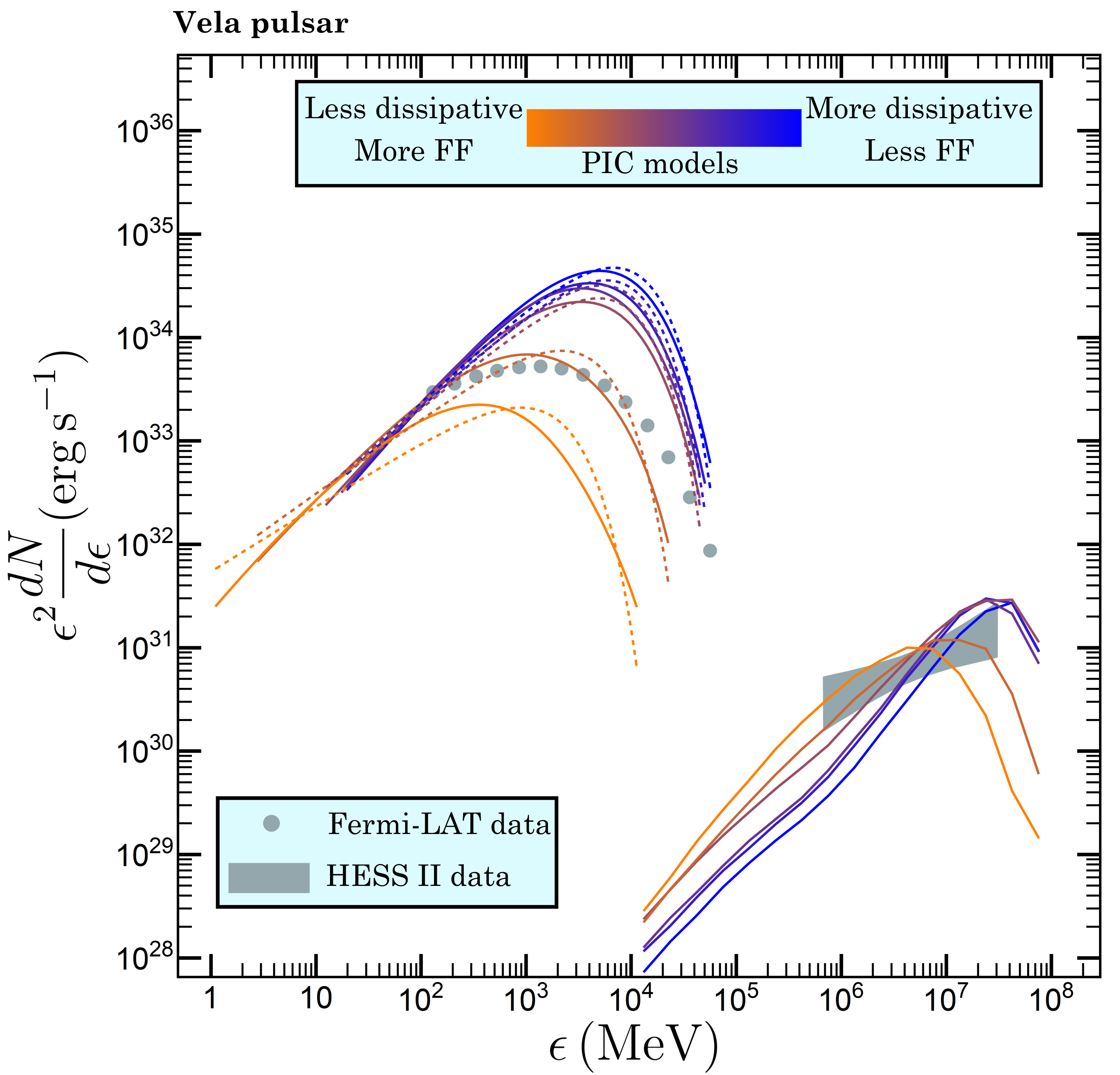}
    \caption{Curvature-radiation and IC spectra from the PIC models of \citet{2023ApJ...954..204K} for $\alpha = 45^{\circ}$, compared with Fermi-LAT (gray dots) and H.E.S.S. II (gray band) data for the Vela pulsar. The PIC curvature spectra peak in the GeV range, while the IC spectra peak at TeV energies. The curve color encodes the degree of FFness: orange denotes more FF (less dissipative) models with higher particle injection rates in the separatrix zone, and blue denotes more dissipative models with lower injection rates (see \citealt{2023ApJ...954..204K} for the corresponding quantitative injection-rate values). Solid curves represent the direct PIC curvature spectra, while dashed curves show best fits (to the PIC spectra) with exponential-cutoff power laws (see \citealt{2023ApJ...954..204K} for details). The IC spectra are computed using PIC particle distributions and a uniform power-law seed-photon field. The model that best matches the Fermi-LAT spectrum also reproduces the observed H.E.S.S. shape, demonstrating a unified GeV-TeV emission framework.}
    \label{fig:vela spectrum}
\end{figure}

We calculate the IC scattering spectrum numerically via a Monte Carlo approach. The differential photon production rate due to IC scattering, expressed in the lab frame, is given by \citep[][see also \citealt{2015ApJ...811...63H,2025A&A...695A..93C}]{1965PhRv..137.1306J,1970RvMP...42..237B}:
\begin{equation}
\frac{dN}{dt d\epsilon_s\,d\epsilon_t} =  \frac{2\pi r_e^2\,m_e\,c^3}{\gamma_{\rm L}^2\epsilon_t}\left[2q_{\epsilon}\ln q_{\epsilon}+(1+2q_{\epsilon})(1-q_{\epsilon})+\frac{1}{2}\frac{(\Gamma_{\epsilon}q_{\epsilon})^2}{1+\Gamma_{\epsilon}q_{\epsilon}}\right]n(\epsilon_t)
\label{eq:ic_differential}
\end{equation}
where $r_e=q_e^2/m_ec^2$ is the classical electron radius, $\epsilon_t$ and $\epsilon_s$ are the target and scatter photon energies in units of $m_e c^2$, $n(\epsilon_t)\equiv dn/d\epsilon_t$ is the differential number density of the soft photons while $\Gamma_{\epsilon}$ and $q_{\epsilon}$ are given by $\Gamma_{\epsilon}=4\gL\epsilon_t$ and $q_{\epsilon}=\epsilon_s/[\Gamma_{\epsilon}(\gL-\epsilon_s)]$, respectively. The expression in Eq.~\ref{eq:ic_differential} assumes an isotropic soft photon gas and relativistic electrons and is exact for both the Thomson ($\Gamma_{\epsilon}\ll 1$) and the Klein-Nishina ($\Gamma_{\epsilon}\gg 1$) regimes under the condition that $\gL\gg 1$.

For the target photons, following \cite{2018ApJ...869L..18H}, we consider an energy distribution following a power-law
\begin{equation}
n(\epsilon_t) \propto \epsilon_t^{-\alpha}, \quad \epsilon_t^{\min} \leq \epsilon_t \leq \epsilon_t^{\max},
\end{equation}
with $\alpha=1$ as a convenient approximation to the observed sub-millimeter to IR-optical spectrum of Vela \citep{2003A&A...406..645S,2011MNRAS.415..867D,2013ApJ...775..101Z,2017ApJ...851L..10M,2019ApJ...885L..10L} (mid-IR points are treated cautiously due to likely contamination). We set $\epsilon_t^{\min} = 0.005$ eV and $\epsilon_t^{\max}=4$ eV to bracket the measured band.

To compute the spectrum of scattered photons, we employ the following Monte Carlo procedure: For each macroparticle in the PIC simulation, characterized by its Lorentz factor $\gL$, we calculate the total probability of scattering during a time interval $\Delta t$ as:
\begin{equation}
P_\mathrm{scatter}(\gL) = \Delta t \int_{\epsilon_t^{\min}}^{\epsilon_t^{\max}} \int_{\epsilon_s^{\min}}^{\epsilon_s^{\max}} \frac{dN}{dt\, d\epsilon_s\,d\epsilon_t}(\gL, \epsilon_t, \epsilon_s)\, d\epsilon_s\, d\epsilon_t,
\label{eq:prob_photon_generation}
\end{equation}
where $\epsilon_t$ is the energy of the seed (target) photon and $\epsilon_s$ is the energy of the scattered photon. The limits of integration are set by $\epsilon_s^{\min} = \epsilon_t^{\min}$ and $\epsilon_s^{\max} = \gL\Gamma_{\epsilon}/(1+\Gamma_{\epsilon})$, where $\Gamma_{\epsilon}$ is the dimensionless boost parameter relevant for the scattering kinematics \citep{1970RvMP...42..237B}.

In principle, using a time step $\Delta t$ comparable to that of the PIC simulation yields tiny scattering probabilities, such that the number of scattered photons remains too low to construct a meaningful spectrum. To overcome this, we adopt a larger effective $\Delta t$, ensuring that $P_{\mathrm{scatter}}(\gL) < 1$ for all particles, while increasing the expected number of scattering events across the sample. For each particle, we then draw a random number $ u \in [0,1] $. If $ u < P_\mathrm{scatter}(\gL) $, we register a scattering event and proceed to sample the scattered photon energy from the corresponding differential distribution.

For electrons that undergo scattering, we determine the scattered photon energy $\epsilon_s$ by constructing and inverting the cumulative probability distribution:
\begin{equation}
f_{\mathrm{cum}}(\epsilon|\gL) = \frac{1}{P_\mathrm{scatter}(\gL)}\int_{\epsilon_s^{\min}}^{\epsilon} \int_{\epsilon_t^{\min}}^{\epsilon_t^{\max}} \frac{dN}{dt\, d\epsilon_s\, d\epsilon_t}(\gL, \epsilon_t, \epsilon_s)\, d\epsilon_t\, d\epsilon_s.
\end{equation}

We sample $\epsilon_s$ by drawing a uniform random number and inverting this cumulative function. The resulting scattered photon energies are recorded and binned to construct the synthetic IC spectrum. Repeating this process across the entire ensemble of electrons/positrons in the PIC simulation yields a statistically robust IC spectrum that can be directly compared with observations. 

In this simplified treatment, we neglect back-reaction IC energy losses on the PIC particles, since the curvature radiation losses dominate. The IC spectrum is thus computed in post-processing from the simulation's six-dimensional particle distribution function, with electron/positrons treated as test particles.  

In \cite{2023ApJ...954..204K}, we explored 12 representative combinations of period ($P$) and surface magnetic field ($B_{\star}$) spanning the parameter space occupied by YPs and MSPs. For the present analysis, we focus on the combination whose spin-down power ($\dot{\mathcal{E}}$), computed using the force-free spin-down relation (Eq.~\ref{eq:spin down power}), is closest to that of the Vela pulsar ($\dot{\mathcal{E}}_{\rm Vela} \simeq 6.9\times 10^{36}~\mathrm{erg\,s^{-1}}$). This case is: $(P = 223.9~\mathrm{ms},\, B_{\star} = 3.98\times 10^{12}~\mathrm{G})$, yielding $\dot{\mathcal{E}} \simeq 10^{37}~\mathrm{erg\,s^{-1}}$. For this parameter set, six different levels of FF-ness\footnote{Here ``FF-ness" parameterizes proximity to the ideal force-free state: larger FF-ness corresponds to higher particle injection in the separatrix zone, more effective screening of the accelerating electric field components, and thus less-dissipative magnetospheres; smaller FF-ness implies lower injection and more dissipative solutions \citep{2023ApJ...954..204K}.} were considered, corresponding to different assumptions about the pair-production efficiency in the separatrix zone.

In Fig.~\ref{fig:vela spectrum}, we compare the observed phase-averaged spectrum of the Vela pulsar from Fermi-LAT and H.E.S.S. II (shown as gray data points and band) to the model spectra obtained from the selected PIC simulation. Colored curves denote different levels of FF-ness, as indicated in the legend. The model curvature radiation spectra are plotted with physically motivated normalizations, directly set by the PIC particle distributions. In contrast, the IC spectra are rescaled in overall amplitude, by adjusting the soft-photon normalization in each case, to match the observed TeV flux levels best. We find that the second-least dissipative model (dark orange curve) reproduces both the shape and normalization of the observed GeV spectrum most accurately. Remarkably, this same model also yields an IC spectrum whose shape closely matches the TeV-band observations from H.E.S.S. II, suggesting a coherent, unified explanation of the GeV and TeV emission within this framework across both energy regimes.

\section{Discussion and Conclusions}\label{sec:Discussion Conclusions}

\subsection{Consolidated Picture and Key Results}\label{Consolidated Picture and Key Results}

We present a unified, observation-anchored framework for gamma-ray pulsars that ties global magnetospheric structure to population phenomenology for their phase-averaged emission. Starting from FF/FIDO insights and 3D PIC simulations, we model the ECS as the principal accelerator/emitter and derive two physically distinct operating branches for curvature radiation: an RRLmax branch and a PDL branch. These define a transition spin-down power, $\dot{\mathcal E}_{\rm TR}$, and a quantified FP luminosity function. We then project ATNF pulsars and McGill magnetars onto both the FP and the $P$-$\dot P$ plane, compute (under a maximal assumption) $\epsilon_{\rm c}$, $L_\gamma$, and $F_\gamma$ for each source, and compare to the  {Fermi}-LAT sample. Finally, we assess the role of physically motivated scaling factors $(\eta_{\RLC},\eta_{\BLC},\eta_{\rm pc},\eta_\alpha)$, identify degeneracies and constraints, show that the data deviate from the RRLmax envelope in a way suggestive of pair-regulated saturation, and demonstrate GeV-TeV consistency for Vela using PIC particle distributions plus a minimal seed-photon model. Finally, we translate these results into concrete observability forecasts, most notably a predicted MeV-dominated, GeV-faint population relevant to next-generation MeV missions.

The key results of this study are:

\begin{enumerate}[label=(\roman*)]
\item \textbf{Two-regime structure and scalings.}
The FP envelope is set by (a) an RRLmax branch with $\epsilon_{\rm c}\propto \dot{\mathcal E}^{7/16}$ and $L_\gamma\propto \dot{\mathcal E}$ at nearly constant efficiency, and (b) a PDL branch in which $\epsilon_{\rm c}\propto \dot{\mathcal E}^{7/4}$ (steep rise) and the efficiency decreases (toward lower $\ed$) accordingly. The transition occurs at $\dot{\mathcal E}_{\rm TR}$, whose scaling follows $\dot{\mathcal E}_{\rm TR}\propto B_\star^{2/7} r_6^{6/7} \eta_{\RLC}^{8/7} \eta_{\BLC}^{4/7} \eta_{\rm pc}^{-16/7} \eta_{\alpha}\propto P^{2/3} \eta_{\RLC}^{4/3} \eta_{\BLC}^{2/3} \eta_{\rm pc}^{-8/3} \eta_{\alpha}$ $\simeq 10^{32}\mathrm{erg\,s^{-1}}$ for YPs, setting a physically motivated “knee’’ in both $\epsilon_{\rm c}$ and $\eta_{\rm eff}$.

\item \textbf{Role of scaling parameters and degeneracies.}
Varying $(\eta_{\RLC},\eta_{\BLC},\eta_{\rm pc})$ shifts the RRLmax envelope and $\dot{\mathcal E}_{\rm TR}$ in predictable ways; different combinations can mimic similar death-line behavior. Nevertheless, the joint information of the two regime slopes and the transition location constrains viable combinations. In practice, PIC outputs are well summarized by \emph{effective} scale factors that capture the global behavior despite underlying particle-population distributions.

\item \textbf{FP normalization constrains the number of emitters.}
Comparing the quantified theoretical FP relation to the observed FP exponents and normalization sets bounds on the product of emitting multiplicity and dissipative volume, $C_{\rm MV}=M_{\rm em}f_V\approx 1.3\, r_6^{-1/2} \, \eta_{\RLC}^{2/3}\, \eta_{\alpha}^{5/12}$. This provides a population-level constraint on the number of particles radiating in the ECS across pulsars.

\item \textbf{Visibility mapping and catalog comparison.}
Mapping $\dot{\mathcal E}_{\rm TR}$ onto the $P$-$\dot P$ plane defines a \emph{gamma-ray visibility line}, i.e., $\dot{P}_{\rm vis}\approx 3\times10^{-15}\,\eta_{\RLC}^{4/3}\eta_{\BLC}^{2/3}\eta_{\rm pc}^{-8/3}\,\eta_{\alpha}\,P^{11/3}$. Applying this to the ATNF+McGill sample (under the maximal $L_{\gamma},\;\epsilon_{\rm c}$ assumption) and comparing with the  {Fermi}-LAT catalog shows the expected population split: MSPs are predominantly on the observable side, YPs cluster around the boundary, and magnetars lie mostly below it. Below the line, the rapid decline of $\epsilon_{\rm c}$ and $F_\gamma$ accounts for the paucity of detections; above it, detectability is set mainly by distance, beaming geometry, and phase-coherent search depth.

\item \textbf{High-$\dot{\mathcal E}$ deviation from the RRLmax envelop and pair-regulated interpretation.}
For both MSPs and YPs, the FP data reveal a clear, systematic departure of the upper $\epsilon_{\rm c}$ envelop from the RRLmax scaling at high $\dot{\mathcal E}$. More specifically, the envelope flattens and falls below the expected trend. This empirical break is a direct outcome of our analysis. In the discussion below, we show that the departure occurs at a common $\epsilon_{\rm c}$ for MSPs and YPs and argue that it is naturally interpreted as the onset of pair-regulated screening in the ECS, qualitatively consistent with PIC results and compactness considerations, although a fully quantitative treatment is left for future work.

\item \textbf{Magnetars and high-B pulsars: predictions and MSP follow-up.}
Because the highest-$\dot{\mathcal E}$ magnetars sit close to the visibility line, modest additional screening (higher pair yield) can push them below detectability, consistent with their non-detections. Conversely, a characteristic subset of high-$\dot{\mathcal E}$, high-$B_\star$ MSPs above the line remains undetected; targeted Fermi-LAT searches are well motivated, provided contemporaneous, phase-connected radio ephemerides are available to enable deeper phase-coherent integrations.

\item \textbf{GeV-TeV unification from the ECS.}
With PIC particle distributions and a minimal seed-photon field, the same ECS population that produces the GeV curvature component also yields a pulsed TeV IC component consistent in shape with Vela, without fine-tuning, supporting a clean interpretation of pulsar high-energy emission across decades in photon energy.

\item \textbf{MeV-bright, GeV-faint predictions.}
The framework predicts a sizeable corridor of MeV-bright, GeV-faint pulsars below current LAT sensitivity—clear targets for next-generation MeV observatories (e.g.,  {AMEGO-X}-class and TPC-based concepts), with sensitivity goals directly inferable from our FP-driven visibility analysis.
\end{enumerate}

\subsection{FP Interpretation: Death Border, Visibility, and Microphysics}\label{sec:FP Interpretation: Death Border, Visibility, and Microphysics}

\begin{figure}
    \centering
    \includegraphics[width=1.0\linewidth]{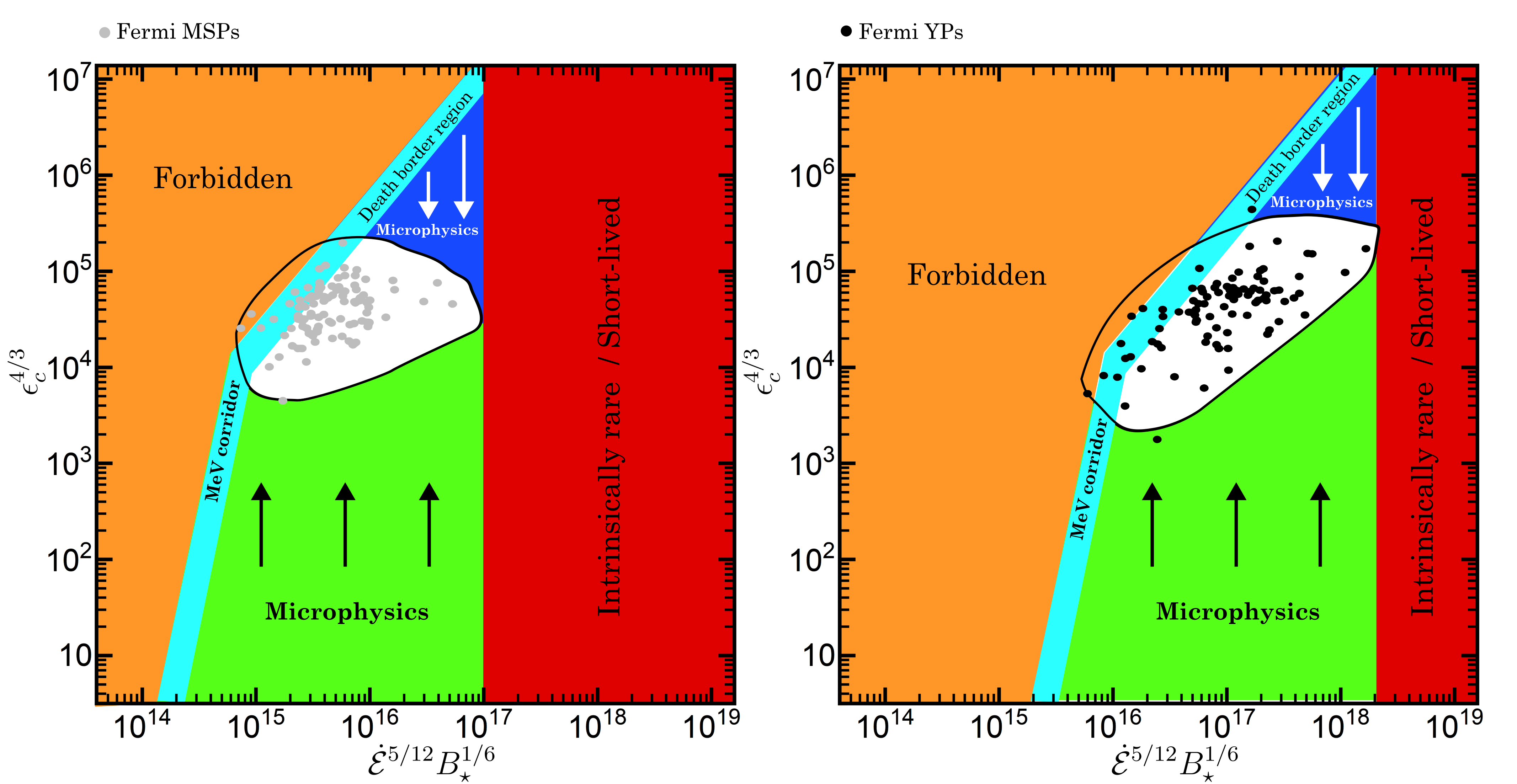}
    \caption{Projection of the FP for MSPs (left) and YPs (right), illustrating the physically distinct regions and their microphysical interpretations. Black contours enclose the observed Fermi domains. Red areas mark intrinsically rare or short-lived populations; orange zones are forbidden by energetic constraints. The cyan band delineates the death border, forming the characteristic knee that separates the maximal RRL regime from the potential-drop-limited regime. Its lower segment marks the \emph{MeV corridor}, where MeV-bright pulsars are predicted to exist but remain challenging to detect with current instruments. The green and blue regions correspond to parameter domains where our global PIC-based models (agnostic to plasma injection microphysics) predict sources without observational counterparts. In the green zones, actual pair production is likely less efficient than in our models, implying weaker screening and higher $\epsilon_{\rm c}$ values. In contrast, in the blue zones, pair production is more efficient, yielding stronger screening and lower $\epsilon_{\rm c}$ values.}
    \label{fig:FP projection schematic}
\end{figure}

To synthesize these results, Fig.~\ref{fig:FP projection schematic} presents a schematic projection of the FP, analogous to the earlier projections: the left-hand panel shows the LAT-detected MSPs (gray points), and the right-hand panel the YPs (black points). Red regions mark parameter combinations expected to be intrinsically rare or short-lived; orange regions are forbidden by energetic and radiative constraints. The narrow cyan band labeled \emph{death border} delineates the boundary of this forbidden domain. This band produces the characteristic knee, which currently acts as \emph{visibility border}, i.e., the observational detection frontier with current GeV sensitivity. Just below this border lies a corridor where MeV-bright, GeV-faint pulsars are expected, despite their low gamma-ray efficiency.

Before turning to the microphysical interpretation encoded by this schematic, we note an observational caveat. A small class of ``soft-gamma/MeV pulsars'' are already known to be bright at hard X-ray to MeV energies while remaining weak or undetected in the GeV band \citep{2015MNRAS.449.3827K}. Within caustic geometries in which the dominant high-energy emission arises in or near the ECS, such sources need not be in tension with the population-level picture developed here or with the broader LAT trends. If the observer line of sight samples primarily the periphery of the high-energy sky map rather than the brightest caustic core, the pulse morphology can appear single-peaked, as is common among the soft-gamma/MeV pulsars, and the observed spectrum can be biased toward lower effective cutoff energies even for relatively high $\dot{\mathcal E}$ \citep{2017ifs..confE...6H}. In our framework, this corresponds to viewing configurations that preferentially sample emission from regions with a smaller effective accelerating field and possibly a different radius of curvature, suppressing the LAT-band component while leaving a bright MeV contribution. At the same time, their concentration at high $\dot{\mathcal E}$ and extreme hard X-ray luminosities leaves open the possibility that their local pair yield and dissipation differ in a systematic way from the bulk LAT population, so geometry may not be the only ingredient. Distinguishing between predominantly geometric versus microphysical explanations will require phase-resolved, high-resolution forward modeling of sky maps and more sensitive MeV instruments that can substantially expand the sample of MeV-dominated pulsars, enabling sharper tests of whether these sources are geometric outliers of the same ECS-curvature sequence and placing stronger constraints on the underlying acceleration and beaming structure. With this geometric caveat in mind, we now interpret what Fig.~\ref{fig:FP projection schematic} implies about pair creation and screening across the broader population. 

The schematic shown in Fig.~\ref{fig:FP projection schematic} encodes constraints on the pair-production microphysics, especially on magnetospheric pair-production efficiency. In the green and blue zones, our PIC-based models allow parameter values that lack detected pulsars. The interpretation is asymmetric: in the green zones, the actual pair yield is likely  {lower} than in our (arbitrarily prescribed) injections, implying weaker screening, larger $\epsilon_{\rm c}$, and an upward shift relative to the model loci; in the blue zones, the actual pair yield is likely higher, implying stronger screening, smaller $\epsilon_{\rm c}$, and a downward shift. The interpretation naturally explains why the Fermi pulsars depart from the RRLmax envelope at higher $\dot{\mathcal E}$.

Although enhanced pair creation can arise in single-photon polar-cap cascades, the abrupt and nearly common departure from the RRLmax branch at similar $\epsilon_{\rm c}$ might be more naturally explained by the activation of an additional site of two-photon pair production, plausibly the ECS \citep{1996A&A...311..172L}, than by a gradual monotonic increase of polar-cap multiplicity alone \citep{2011ApJ...726L..10H,2011ApJ...743..181H,2015ApJ...810..144T,2019ApJ...871...12T}. This is especially striking because the departure appears in both YPs and MSPs, even though polar-cap pair cascades operate in quite different dynamical regimes in the two classes, with MSPs likely in the RRL regime near the polar caps \citep[e.g. Appendix B in][]{2015ApJ...810..144T}.   

For two-photon pair creation, the Breit-Wheeler cross section for $\gamma\gamma\rightarrow e^\pm$ peaks at $\sigma_{\gamma\gamma} \sim 0.3\sigma_T$ when the two photon energies $\epsilon_{1,2}$ slightly exceed the threshold  $\epsilon_1\epsilon_2 (1-\cos\theta_{12}) = 2 (m_e c^2)^2$ where $\theta_{12}$ is the angle of the collision. A delta-function approximation of the cross section captures well the relevant physics and scales for nearly head-on collisions \citep{1985ApJ...294L..79Z}. This motivates characterizing the efficacy of ECS pair production in terms of an effective photon column, i.e., a compactness parameter.

To appraise the efficacy of an additional two-photon pair-production site in the ECS, we introduce a compactness parameter \citep{2009herb.book.....D}, which quantifies the photon-energy column independently of the escape geometry. We define
\begin{equation}
\mathcal{C}
\;\equiv\;
\frac{u_\gamma\,\sigma_{\rm T}\,\lambda_{\rm eff}}{m_e c^2}
\;=\;
\frac{L_\gamma}{A_{\rm esc} c}\,
\frac{\sigma_{\rm T}\,\lambda_{\rm eff}}{m_e c^2}
\;=\;
\frac{\sigma_{\rm T}\,L_\gamma}{m_e c^3}\,
\frac{\lambda_{\rm eff}}{A_{\rm esc}}\,,
\label{eq:C_general}
\end{equation}
where $u_\gamma$ is the local gamma-ray energy density, $\lambda_{\rm eff}$ is the effective length scale to escape along the photon direction within the emitting zone, and $A_{\rm esc}$ is the area through which those photons ultimately exit. Thus $\mathcal{C}$ is the photon energy column $u_\gamma\lambda_{\rm eff}$ expressed in units of $m_ec^2/\sigma_{\rm T}$.

In the pulsar context, we further consider $\lambda_{\rm eff}=\eta_\lambda\,R_{\rm LC}$ and $A_{\rm esc}=\eta_A\,R_{\rm LC}^2$, with $\eta_\lambda$ and $\eta_A$ set by the emission/escape geometry (photon beaming and ECS topology). Substituting into Eq.~\eqref{eq:C_general} gives
\begin{equation}
\mathcal{C}
=\frac{\sigma_{\rm T}\,L_\gamma}{m_e c^3}\,
\frac{\eta_\lambda}{\eta_A}\,
\frac{1}{R_{\rm LC}}
\equiv
\frac{\sigma_{\rm T}\,L_\gamma}{m_e c^3}\,
\frac{1}{\eta_{\rm gm}\,R_{\rm LC}}\,,\qquad
\eta_{\rm gm}\equiv\frac{\eta_A}{\eta_\lambda}\,,
\label{eq:C_geom_factor}
\end{equation}
so that all geometric dependence is absorbed into the single factor $\eta_{\rm gm}$ (the ratio of escape area to path length in units of $R_{\rm LC}$ and $R_{\rm LC}^2$). This formulation is agnostic to whether photons exit through faces or edges of the emitting area.

With the geometry-agnostic compactness of Eq.~\eqref{eq:C_geom_factor}, the $\gamma\gamma$ optical depth at energy $\epsilon$ reads
\begin{equation}
\tau_{\gamma\gamma}(\epsilon)=\eta_s\,\mathcal{K}_{\gamma\gamma\rightarrow e^\pm}(\epsilon)\,\mathcal{C},
\end{equation}
where $\eta_{\rm s}$ is the local soft-to-$\gamma$ luminosity ratio and $\mathcal{K}_{\gamma\gamma\rightarrow e^\pm}(\epsilon)$ encodes the energy/angle dependence of the $\gamma\gamma$ kernel (peaking near threshold and decreasing far above threshold).
The condition $\tau_{\gamma\gamma}\gtrsim 1$ is a spectral (opacity) threshold that would imprint strong attenuation in the GeV band. 
Nonetheless, the FP phenomenology does not require such attenuation. 
The energization and screening of the ECS are governed by a dynamical activation threshold, where pair production begins to influence the accelerating electric field, $E_{\rm acc}$, even in the optically thin regime, i.e., $\tau_{\gamma\gamma}\ll 1$.

In the optically thin regime, a fraction $\tau_{\gamma\gamma}$ of the GeV photons converts, and for GeV-on-keV collisions, each absorbed GeV photon produces one $e^+e^-$ pair with Lorentz factor $\sim10^3$. Hence, the pair injection rate is\footnote{The corresponding pair rest-mass luminosity is 
$L_{\rm rest}=2m_e c^2\,\dot N_{\gamma\gamma\rightarrow e^\pm}=(2m_e c^2/\epsilon_{\rm c})\,\tau_{\gamma\gamma}\,L_\gamma$, which for $\epsilon_{\rm c}=1~{\rm GeV}$ gives
$L_{\rm rest}\approx 10^{-3}\,\tau_{\gamma\gamma}L_\gamma$.}
\begin{equation}
\dot N_{\gamma\gamma\rightarrow e^\pm}\;\simeq\;\tau_{\gamma\gamma}(\epsilon)\,\frac{L_\gamma}{\epsilon_{\rm c}}
\;=\;\eta_s\,\mathcal{K}_{\gamma\gamma\rightarrow e^\pm}(\epsilon)\,\mathcal{C}\,\frac{L_\gamma}{\epsilon_{\rm c}}.
\end{equation}
Defining the onset of effective pair production by the condition $\dot N_{\gamma\gamma\rightarrow e^\pm}\gtrsim\kappa_{\rm req}\,\dot N_{\rm GJ}^{\rm (PC)}$, where
$\kappa_{\rm req}$ denotes the required effective multiplicity and $\dot N_{\rm GJ}^{\rm (PC)}=4\pi^2 B_\star r_\star^3/(q_e c P^2)$ represents the Goldreich-Julian particle outflow from the polar cap, we obtain, using Eqs.~\ref{eq:spin down power} and \ref{eq:C_geom_factor}, the corresponding condition for $\epsilon_{\rm c}$, which reads
\begin{equation}
\epsilon_{\rm c-thr}=
\frac{\sigma_{\rm T}\,q_e\,\eta_\alpha^{1/4}}{m_e c^{15/4}\,r_\star^{3/2}}
\frac{\eta_s\,\mathcal{K}_{\gamma\gamma\rightarrow e^\pm}}{\eta_{\rm gm}\,\kappa_{\rm req}}\;
\frac{L_\gamma^{2}}{B_\star^{1/2}\,\dot{\mathcal E}^{1/4}},\qquad \text{with }\epsilon_{\rm c}\lesssim\epsilon_{\rm c-thr}.
\label{eq:ecut_threshold}
\end{equation}
Equation~\ref{eq:ecut_threshold} expresses the threshold cutoff energy, $\epsilon_{\rm c-thr}$, as a function of the accelerator observables $L_\gamma$, $B_\star$, and $\dot{\mathcal E}$, multiplied by a geometry/feedback factor $\eta_s\,\mathcal{K}_{\gamma\gamma\rightarrow e^\pm}/(\eta_{\rm gm}\,\kappa_{\rm req})$. The factors $\eta_{\rm gm}$ and $\kappa_{\rm req}$, set by geometry, should not vary strongly with $\dot{\mathcal E}$. If, in addition, $\eta_s\,\mathcal{K}_{\gamma\gamma\rightarrow e^\pm}$ varies only weakly across the population below the deviation point (Fig.~\ref{fig:FP projected Fermi PIC Theory - moving average trend}), then $\epsilon_{\rm c-thr}$ effectively depends only on $L_\gamma$, $B_\star$, and $\dot{\mathcal E}$. A positive feedback is also expected: pair-fed synchrotron and synchrotron self-Compton (SSC) emission within the ECS increases the local soft-photon fraction $\eta_s$ and broadens the spectral overlap with the GeV component, consistent with the \citetalias{Smith23} empirical trend that higher-$\dot{\mathcal E}$ pulsars exhibit broader gamma-ray SEDs.

In Fig.~\ref{fig:FP projection ecthreshold} we plot the corresponding $\epsilon_{\rm c-thr}$ loci on the projected FP for $B_\star=10^8$\,G (red; MSPs) and $B_\star=10^{12}$\,G (blue; YPs), adopting a unit modulating factor\footnote{The modulating factor reflects geometry, angular distributions, current-closure constraints, and the local soft-photon SED, and may vary across the population. Under plausible ECS conditions the combined factor is plausibly of order unity, but a full quantification is deferred to dedicated advanced modeling.}, $\eta_s\,\mathcal{K}_{\gamma\gamma\rightarrow e^\pm}/(\eta_{\rm gm}\,\kappa_{\rm req})\sim 1$. We also consider the $L_\gamma$ expressions from Eqs.~\ref{eq:Lgamma rrx} and \ref{eq:Lgamma maxdrop}. For reference, we overlay the right-hand panel of Fig.~\ref{fig:FP projected Fermi PIC Theory - moving average trend} with high transparency. The crossing where $\epsilon_{\rm c}<\epsilon_{\rm c-thr}$ occurs at nearly the same $\epsilon_{\rm c}$ for both populations and coincides with the empirically inferred inflection point, consistent with a common compactness activation in the ECS that enhances pair production (and thus screening) as the RRLmax branch turns over.

Our geometry-agnostic compactness and the dynamical activation threshold in $\epsilon_{\rm c}$ formalize, at the population level, the optically thin current-sheet pair feedback proposed by \citet{2019ApJ...877...53H}. Using local PIC simulations, they showed that even when most photons leave the upstream region unaffected, reconnection in the ECS can self-supply a co-spatial soft field and inject pairs that self-regulate the accelerator. Our suggestion extends this local picture to population data by providing a predictive condition for the cutoff-energy crossing $\epsilon_{\rm c}\simeq \epsilon_{\rm c-thr}$ on the FP and explains why MSPs and YPs depart from the maximal-RRL branch at similar $\epsilon_{\rm c}$ without requiring GeV-band attenuation.

It is important to distinguish both the interaction channel and the criterion being tested. Our analysis concerns the GeV/keV channel in or near the ECS and focuses on a dynamical activation threshold (pair feedback), rather than spectral attenuation. In contrast, \citet{2021ApJ...923..194H} modeled the very high-energy (VHE), up to multi-TeV, IC scattering component interacting with optical-IR targets and found strong $\gamma\gamma$ attenuation for Crab but not for Vela, i.e., $\tau_{\gamma\gamma}\gg 1$ for Crab and $\tau_{\gamma\gamma} \ll 1$ for Vela. The \citet{2021ApJ...923..194H} results imply a small $\eta_s\mathcal{K}_{\gamma\gamma\rightarrow e^\pm}$ in the VHE/optical-IR channel for some sources, but they do not constrain the GeV/keV regime considered here, where the target photon energies and angular factors differ and where activation can occur even for $\tau_{\gamma\gamma}\ll 1$. Our geometry-agnostic compactness and the associated activation threshold therefore complement, rather than contradict, the source-specific attenuation constraints derived for the VHE band.

The compactness-based activation framework is intentionally minimal and population-level; it provides a necessary condition for ECS pair feedback and naturally captures the near-common cutoff energy at which MSPs and YPs depart from the RRLmax branch, but it is not a proof of sufficiency. Demonstrating that this mechanism operates in specific sources will require global, time-dependent kinetic radiation modeling that self-consistently couples reconnection-driven particle acceleration, anisotropic soft-photon production (synchrotron/SSC, secondary pair synchrotron, and thermal keV fields), full-angle $\gamma\gamma$ transport with relativistic aberration, and pair back-reaction on $E_{\rm acc}$. Such calculations are needed to determine $\eta_s(\dot{\mathcal E})$, evaluate the effective kernel $\mathcal{K}_{\gamma\gamma\rightarrow e^\pm}(\epsilon)$ in the ECS geometry, constrain the geometry/current factors $\eta_{\rm gm}$ and $\kappa_{\rm req}$, and calibrate their dispersion against source-resolved SEDs (e.g., Crab, Vela, bright MSPs). Until then, we regard $\mathcal{C}$ and $\epsilon_{\rm c-thr}$ as a physically motivated, testable organizing principle for the observed FP behavior, pending advanced modeling and targeted source-level validation.

\subsection{Prospects for Pulsed High Energy Neutrinos From $\gamma\gamma \rightarrow \mu^{\pm}$}

The same compactness-based activation framework can also induce the leptonic channel
$\gamma\gamma \rightarrow \mu^{\pm}$ in or near the ECS of high-$\dot{\mathcal E}$ pulsars. The kinematic threshold $\epsilon_1\,\epsilon_2\,(1-\cos\theta_{12})\ \ge\ 2\,m_\mu^{2}c^{4}$ indicates that $\gtrsim$MeV targets can trigger muon production with GeV photons in near head-on interactions.

Similarly to before, we write the optical depth as
\begin{equation}
\tau_{\mu\mu}(\epsilon)\;=\;\eta_s^{(\mathrm{MeV})}\,\mathcal{K}_{\gamma\gamma\rightarrow \mu^\pm}(\epsilon)\,\mathcal{C},
\end{equation}
where $\eta_s$ is now the luminosity ratio between $\sim$MeV and GeV and $\mathcal{K_{\gamma\gamma\rightarrow \mu^\pm}}$ is the corresponding kernel for muons. Thus, the muon-pair injection rate is
\begin{equation}
\dot N_{\gamma\gamma \rightarrow \mu^{\pm}}\ \simeq\ \eta_s^{(\mathrm{MeV})}\,\mathcal{K}_{\gamma\gamma\rightarrow \mu^\pm}(\epsilon)\,\mathcal{C}\,\frac{L_\gamma}{\epsilon_{\rm c}}.
\end{equation}
Assuming identical angular distributions and matched soft-SED shapes (normalized over their respective threshold bands) for the $e^\pm$ and $\mu^\pm$ channels, the kernel rescales purely by the lepton mass ratio (the $\mu^\pm/e^\pm$ production cross section ratio), i.e., $\mathcal{K}_{\gamma\gamma\rightarrow \mu^\pm}=(m_e/m_\mu)^{2}\mathcal{K}_{\gamma\gamma\rightarrow e^\pm}$ and so\footnote{Departures from identical angle/SED weighting introduce only order-unity corrections.}
\begin{equation}
\frac{\dot N_{\mu\mu}}{\dot N_{ee}}\;\simeq\;
\Big(\frac{m_e}{m_\mu}\Big)^{\!2}\,
\frac{\eta_s^{(\mathrm{MeV})}}{\eta_s^{(\mathrm{keV})}}\,,
\end{equation}
where the $\eta_s$ defined in Section~\ref{sec:FP Interpretation: Death Border, Visibility, and Microphysics} is denoted with a superscript ``keV" for clarity. These muons are spawned at mildly relativistic Lorentz factors.

Subsequently, muons are further accelerated in the ECS to an average Lorentz factor $\gamma_{\rm L\mu}$ (prior to decay). The power carried by the accelerated muon population is then
\begin{equation}
L_\mu \;\simeq\; 2\,\gamma_{\rm L\mu}\,m_\mu c^2\,\dot N_{\mu\mu}.
\end{equation}
Muon decays proceed primarily \citep{ParticleDataGroup:2024cfk} via  
\begin{equation}
\begin{aligned}
\mu^{+} &\to e^{+} + \nu_{e} + \bar{\nu}_{\mu},\\
\mu^{-} &\to e^{-} + \bar{\nu}_{e} + \nu_{\mu}.
\end{aligned}
\end{equation}
The decay kinematics imply that, on average, the two neutrinos take $\simeq 2/3$ of the muon energy, with characteristic mean per neutrino
$\epsilon_\nu \approx (1/3)\,\gamma_{\rm L\mu} m_\mu c^2$.
Thus, the \emph{pulsed} neutrino luminosity is
\begin{equation}
L_\nu \ \simeq\ \frac{2}{3}\,L_\mu
\ =\ \frac{4}{3}\,\gamma_{\rm L\mu}\,m_\mu c^2\,
\eta_s^{(\mathrm{MeV})}\,\mathcal{K}_{\gamma\gamma\rightarrow \mu^\pm}(\epsilon)\,\mathcal{C}\,\frac{L_\gamma}{\epsilon_{\rm c}},
\label{eq:Lnu_1}
\end{equation}
and the neutrino light curve should be approximately phased-aligned with the GeV/TeV $\gamma$-ray peaks.

Assuming muons experience the same available potential drop as the $e^{\pm}$ and that synchrotron losses are negligible (scaling as $1/m^2$), while the curvature-reaction maximum, i.e., the RRLmax regime for muons, would require a much larger potential than available\footnote{For muons, the RRL $\gamma_{\rm L}$ is the same as for $e^{\pm}$, but the corresponding energy is higher by a factor $m_\mu/m_e$.}, which means that the muons operate always in the PDL regime. Considering that the muons experience a fraction $\eta_{\rm pdm}$ of the available potential drop (see Equation~\ref{eq:gammaL max}), we get
\begin{equation}
\gamma_{\rm L\mu} \simeq \frac{\eta_{\rm pdm}\,4\pi^2 q_e r_\star^3 B_\star}{m_\mu c^4\, P^2}\,.
\end{equation}
For representative LAT pulsars, the corresponding ECS acceleration time to $\gamma_{\rm L\mu}$, i.e., $t_{\rm acc}^\mu=\gamma_{\rm L\mu} m_\mu c/(q_e\, \eta_{B_{\rm LC}} B_{\rm LC})$, is shorter than the lab-frame muon decay time, $\gamma_{\rm L\mu}\tau_\mu$ (with $\tau_\mu\simeq 2.2~\mu{\rm s}$), so a substantial fraction of muons can be accelerated before decaying and the resulting neutrino signal remains phase-coherent with the GeV/TeV gamma-ray emission. Substituting $\gamma_{\rm L\mu}$ into the Eq.~\ref{eq:Lnu_1}, we get
\begin{equation}
\frac{L_\nu}{L_\gamma} \simeq \eta_{\rm pdm}\frac{16\pi^2 q_e r_\star^3 B_\star}{3c^2\, P^2\,\epsilon_{\rm c}}\,
\eta_s^{(\mathrm{MeV})}\,\Big(\frac{m_e}{m_\mu}\Big)^{2}\,\mathcal{K}_{\gamma\gamma\rightarrow e^\pm}(\epsilon)\,\mathcal{C},
\label{eq:Lnu/Lgamma 1}
\end{equation}
where we have also used that $\mathcal{K}_{\gamma\gamma\rightarrow \mu^\pm}=(m_e/m_\mu)^2\mathcal{K}_{\gamma\gamma\rightarrow e^\pm}$.

Invoking the $e^\pm$ activation condition $\eta_s^{(\mathrm{keV})}\mathcal{K}_{\gamma\gamma\rightarrow e^\pm}/(\eta_{\rm gm}\kappa_{\rm req})\simeq 1$ implied in Section~\ref{sec:FP Interpretation: Death Border, Visibility, and Microphysics} and taking into account Eq.~\ref{eq:C_geom_factor}, we get 
\begin{equation}
\frac{L_\nu}{L_\gamma} \simeq \eta_{\rm pdm}\frac{16\pi^2 q_e r_\star^3 B_\star}{3c^2\, P^2\,\epsilon_{\rm c}}\,
\Big(\frac{m_e}{m_\mu}\Big)^{2}\,f_{\rm Mk}\,\kappa_{\rm req}\,\frac{\sigma_T L_\gamma}{m_e c^3 R_{\rm LC}}\,,
\label{eq:Lnu/Lgamma 2}
\end{equation}
where $f_{\rm Mk}= \eta_s^{(\mathrm{MeV})}/\eta_s^{(\mathrm{keV})}$.

\begin{figure}
    \centering
    \includegraphics[width=0.5\linewidth]{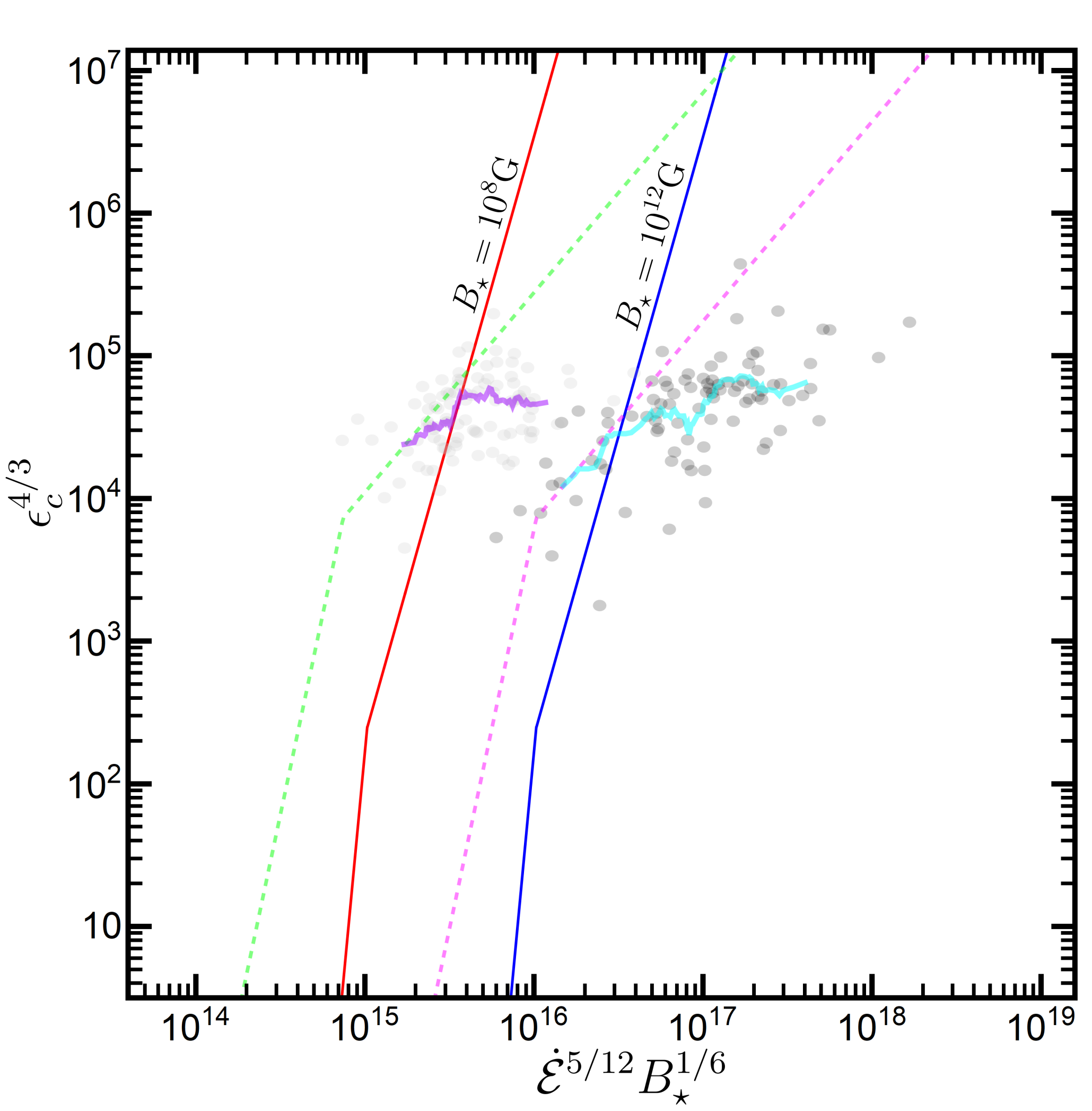}
    \caption{Loci of $\epsilon_{\rm c\text{-}thr}$ on the projected FP for $B_\star=10^{8}\,\mathrm{G}$ (red line; MSPs) and $B_\star=10^{12}\,\mathrm{G}$ (blue line; YPs). For reference, the right-hand panel of Fig.~\ref{fig:FP projected Fermi PIC Theory - moving average trend} is overplotted with high transparency. The pair-activation criterion $\epsilon_{\rm c}\lesssim\epsilon_{\rm c\text{-}thr}$ is satisfied at nearly the same $\epsilon_{\rm c}$ for both populations and near the observed departure from the maximal-RRL branch. This alignment, tentative but suggestive, indicates that enhanced pair creation, and the attendant screening of $E_{\rm acc}$, emerge at this transition, with $\eta_{B_{\rm LC}}$ effectively acquiring a dependence on $\dot{\mathcal E}$.}
    \label{fig:FP projection ecthreshold}
\end{figure}

It follows from Eqs.~\ref{eq:Lnu/Lgamma 1} and \ref{eq:Lnu/Lgamma 2} that the ratio $L_\nu/L_\gamma$ is controlled by the available potential drop (through $\gamma_{\rm L\mu}$) and by the effective $\gamma\gamma$ optical depth for $\mu$ production $\eta_s^{(\mathrm{MeV})}\mathcal{K}_{\gamma\gamma\rightarrow e^\pm}\mathcal{C}$; equivalently, adopting the activation condition inferred in Fig.~\ref{fig:FP projection ecthreshold}, by the product $f_{\rm Mk}\kappa_{\rm req}$. For a representative young pulsar near the onset of enhanced $\gamma\gamma$ pair activation ($B_\star=5\times10^{11}\,\mathrm{G}$, $P=0.1\,\mathrm{s}$, $\dot{\mathcal{E}}\approx10^{35}\,\mathrm{erg\,s^{-1}}$), taking $\eta_{\rm pdm}\simeq 0.1$ yields muons reaching $\simeq 60$~TeV and decay neutrinos up to $\simeq 20$~TeV. For $L_\gamma\approx10^{34}\,\mathrm{erg\,s^{-1}}$ this gives $L_\nu/L_\gamma=10^{-3}\,f_{\rm Mk}\kappa_{\rm req}$; if $f_{\rm Mk}\kappa_{\rm req}\gtrsim 1$ then $L_\nu\approx 10^{31}\,\mathrm{erg\,s^{-1}}$. At higher $\dot{\mathcal{E}}$, the enhanced $e^\pm$ creation reduces $\eta_{B_{\rm LC}}$ and hence the realized potential drop, i.e., smaller $\eta_{\rm pdm}$. Using Crab-like parameters and $\eta_{\rm pdm}\simeq 10^{-3}$ \citep[][]{2017ApJ...842...80K,2022ApJ...934...65K}, one obtains $\epsilon_\nu\approx 15$~TeV and $L_\nu/L_\gamma\approx 10^{-2}\,f_{\rm Mk}\kappa_{\rm req}$, implying $L_\nu\gtrsim 10^{34}\,\mathrm{erg\,s^{-1}}$. 

Alternatively, if a substantial population ($\sim10^3$) of energetic MSPs resides in the Galactic disk and bulge, their cumulative $\gamma\gamma\rightarrow\mu^\pm$ output could contribute to the unresolved (quasi-diffuse) Galactic neutrino emission reported by IceCube \citep{2023Sci...380.1338I}. Importantly, the measured Galactic-plane signal is broadly consistent\footnote{However, IceCube notes that a simple GeV-to-100 TeV extrapolation of a gamma-ray–inferred diffuse template underpredicts their best-fitting neutrino flux by a factor $\sim 5$ plausibly reflecting CR-transport/spectral differences toward the inner Galaxy and/or additional unresolved-source contributions.}  with diffuse neutrino production from cosmic-ray interactions in the interstellar medium, so such an MSP contribution is not required and is likely sub-dominant; nevertheless, it could still be non-negligible and would be interesting to quantify with a dedicated population-synthesis calculation \citep[e.g.,][]{2025ApJ...988...78T,2025arXiv251015661S,gammaPTA}. MSPs have also been widely discussed as a possible contributor to the Fermi-LAT Galactic-center GeV excess, although its origin remains debated \citep[e.g.,][]{2011JCAP...03..010A,2011PhRvD..84l3005H,2016JCAP...03..049H,2020JCAP...12..035P,Calore2015,2024PhRvD.109l3042M,2025arXiv250717804L}. In that context, constraints on (or a detection of) a neutrino component from a bulge MSP population would provide a valuable, independent handle on the viability of such scenarios. 

These estimates are illustrative, grounded in a compactness-based activation scheme that provides viable interpretations of the observed phenomenology, and they define concrete, testable targets for future observations and modeling. Future observations, together with more advanced, source-resolved kinetic radiation modeling, will further constrain, and may ultimately rule out, different parameter regimes. In the near term, phase-coherent techniques using known pulsar ephemerides to suppress atmospheric and diffuse backgrounds \citep[e.g.,][]{1989A&A...221..180D,2011ApJ...732...38K,2021ApJ...911...45L} might reach sensitivities to pulsed neutrino luminosities of $L_\nu \lesssim 10^{34}\,\mathrm{erg\,s^{-1}}$ below $50$~TeV for expected bright targets (e.g., Crab) with IceCube and forthcoming facilities.

\subsection{Synchrotron Narratives in the ECS}\label{Synchrotron Narratives in the ECS}

A parallel line of studies has argued that the GeV component in Fermi-detected pulsars arises predominantly from synchrotron radiation in the ECS \citep[e.g., ][]{2016MNRAS.457.2401C,2018ApJ...855...94P,2023ApJ...943..105H}, implicitly assuming emission in fields of order $B_{\rm LC}$. In that case, GeV photons can be produced with $\gamma_{\rm L}\sim 10^{5}$-$10^{6}$ by reconnection alone (with the ECS magnetization as the typical limit for particle acceleration). However, such Lorentz factors are insufficient to account for the recently detected pulsed emission up to $\sim 20$~TeV from Vela via IC scattering: the TeV observations require $\gamma_{\rm L}\gtrsim 10^{7.4}$, thereby implying particle energies well above those typically invoked in a $B_{\rm LC}$-synchrotron scenario. In this sense, the TeV detection provides a decisive constraint: the emitting pairs must regularly reach $\gamma_{\rm L}\gtrsim 10^{7}$.

In response, a branch of this literature has refined the synchrotron picture by appealing to reconnection locales with effectively small $B_\perp\ll\BLC$ (near X-points) and by parameterizing the outcome with the sheet magnetization $\sigma_{\rm M}$ (as a cap on attainable energy) and a synchrotron burnoff Lorentz factor $\gamma_{\rm syn}$, as the cooling-limited scale \citep[][]{2023ApJ...959..122C}. In that language, the regime is controlled by the ratio $\sigma_{\rm M}/\gamma_{\rm syn}$, and GeV cutoffs can be reached when $\sigma_{\rm M} \gg \gamma_{\rm syn}$ because particles experience acceleration in regions of reduced $B_\perp$. This viewpoint has gradually converged toward the FP-based picture. The two-scale competition we identified, between the curvature RRLmax, $\gamma_{\rm L}^{\rm RRLmax}$, and the PDL, $\gamma_{\rm L}^{\rm PDL}$, is directly analogous to their $(\gamma_{\rm syn},\sigma_{\rm M})$. The crossover condition $\sigma_{\rm M}\sim\gamma_{\rm syn}$ maps one-to-one onto our $\gamma_{\rm L}^{\rm PDL}=\gamma_{\rm L}^{\rm RRLmax}$ transition at $\dot{\mathcal E}_{\rm TR}$.

We note that, in practice, the effective $B_\perp$ sampled by the radiating particles and the dwell time spent in low-$B_\perp$ versus radiating regions control $\epsilon_{\rm c}$. Absent a predictive mapping from global parameters to these local fractions, the degree of $B_\perp$ reduction below $B_{\rm LC}$ remains model-dependent, which complicates population-level predictions compared to the curvature-ECS scalings used here.

Our framework retains this two-scale physics but anchors it to global, observable quantities. In the ECS beyond the light cylinder, the relevant curvature radius is set geometrically, $R_c \sim \eta_{\rm RLC}R_{\rm LC}$, so curvature emission naturally accounts for the observed GeV cutoffs with $\gamma_{\rm L} \gtrsim 10^{7}$, and the same particle population produces the TeV component via IC scattering without further assumptions. Moreover, the curvature-ECS picture, combined with a GJ scaling of the emitter, reproduces the empirical FP exponents and yields population-level predictions (RRLmax vs.\ PDL branches, the transition at $\dot{\mathcal E}_{\rm TR}$, and the associated visibility mapping). Beyond reproducing the FP, our framework provides explicit, testable rules for how the observables scale with spin-down power in the different regimes. 

Through the “dictionary’’ discussed above, elements of these predictions could, in principle, be recast within the reconnection-synchrotron paradigm: their $(\sigma_{\rm M},\gamma_{\rm syn})$ pair is the local analogue of our $(\gamma_{\rm L}^{\rm PDL},\gamma_{\rm L}^{\rm RRLmax})$, and the regime selector $\sigma_{\rm M}/\gamma_{\rm syn}$ mirrors our $\gamma_{\rm L}^{\rm PDL}/\gamma_{\rm L}^{\rm RRLmax}$. However, turning that local description into the  {same} FP exponents and the  {same} $\dot{\mathcal E}$-scalings for $\ec$ and $L_\gamma$ requires a specific, population-wide mapping of $\sigma_{\rm M}$ and $\gamma_{\rm syn}$ to the global pulsar parameters ($\dot{\mathcal E},B_\star,P,\alpha$). In our curvature-ECS formulation, those scalings follow directly from the light-cylinder geometry and GJ normalization, hence the predictive power at the catalog level; in the synchrotron picture, the corresponding global closure has not yet been demonstrated and would need to recover, without fine-tuning, $\ec \propto \dot{\mathcal E}^{7/16}$ in the RRLmax branch and the steeper $\propto \dot{\mathcal E}^{7/4}$ decline in the potential-limited branch, together with the observed FP exponents.

A second practical advantage is how naturally the catalogs fit into the global picture. Projecting ATNF pulsars and McGill magnetars onto the FP and onto the $P$-$\dot P$ plane using our scalings reproduces the observed occupancy: MSPs lie predominantly on the observable side of the visibility border, YPs straddle it, and magnetars fall largely below it, with the predicted flux/cutoff behavior tracking the  {Fermi}-LAT detections. While a reconnection-synchrotron treatment could be made compatible via the equivalence noted above, it would similarly need to establish how $\sigma_{\rm M}$ and $\gamma_{\rm syn}$ co-vary with ($\dot{\mathcal E},B_\star,P$)  {across the population} so that the same visibility mapping and detection fractions emerge. In short, the two narrations share the same underlying two-scale competition; our curvature-ECS implementation ties those scales to global observables in a way that yields simple $\dot{\mathcal E}$-laws and catalog-level predictions that can be, and have been, confronted directly with the data.

It is worth emphasizing that the interpretation advanced by \citet{2023ApJ...959..122C} effectively requires large magnetization at the light cylinder, $\sigma_{\rm M}> 10^7$, in the ECS. In practice, this points to low local pair multiplicity in the emitting layer, a potentially challenging requirement for energetic pulsars such as Vela, where polar-cap pair production is expected to be substantial \citep[e.g.,][]{2013MNRAS.429...20T,2015ApJ...810..144T,2019ApJ...871...12T}. More generally, if the global pair yields were uniformly low, it would be challenging to sustain FF-like magnetospheres whose large-scale geometry successfully reproduces the observed gamma-ray light-curve phenomenology, \citep[e.g., ][]{2014ApJ...793...97K}.

A natural reconciliation, consistent with our studies \citep{2017ApJ...842...80K,2018ApJ...857...44K,2019ApJ...883L...4K,2022ApJ...934...65K,2023ApJ...954..204K}, is that pair loading is spatially inhomogeneous. The overall polar-cap cascade can be high, while the multiplicity specifically along the last-open field lines that feed the ECS remains only moderate. This picture aligns with the separatrix-zone (SZ) injection framework of \citet{2023ApJ...954..204K}, in which the gamma-ray luminosity and cutoff are regulated by the particle injection rate into the SZ/ECS, whereas the remainder of the open field-line bundle can remain heavily pair-loaded to uphold a quasi-force-free global structure. In that sense, the high-$\sigma_{\rm M}$ requirement in the emitting sheet (low local multiplicity) is compatible with a globally pair-rich magnetosphere, and it maps cleanly onto the role played by the SZ injection rate in our curvature-ECS paradigm. However, this solution could run counter to the abundant local pair creation in the ECS required to moderate gamma-ray luminosity in high-spin-down-power pulsars like the Crab \citep{2019ApJ...877...53H}.

A second strand within the synchrotron school keeps the GeV component as synchrotron emitted in a field of order $B_{\rm LC}$ but posits {continued} acceleration for many light-cylinder radii so that the same particles subsequently reach $\gamma_{\rm L} \gtrsim 4\times10^7$ and generate the TeV component by IC in the outer wind \citep[][]{2025A&A...695A..93C}. This update alleviates the Vela constraint only by introducing new tensions: (i) if the LC energization is capped by the local magnetization, $\sigma_{\rm M}$, a monotonic gain by a factor $\gtrsim 40$, at $r \gg R_{\rm LC}$, must exceed that cap and thus relies on a specific, quantitatively characterized mechanism for sustained wind-zone acceleration, not yet developed in detail in this framework; (ii) with $B_\perp \propto r^{-1}$, particles at $\gamma_{\rm L} \gtrsim 10^7$ radiate appreciable synchrotron power and retain synchrotron cutoffs $\propto\gamma_{\rm L}^2 B_\perp$ out to tens-hundreds of $R_{\rm LC}$, which would generally produce an additional substantial synchrotron component that is not observed unless $B_\perp$ (or the pitch angle) is kept extremely small along the entire path, in which case the scenario effectively approaches large-$R_c$ motion where the practical distinction from curvature becomes again semantic; (iii) IC luminosity gains premised on an isotropic target (e.g., the CMB) are least realistic on LC scales, while anisotropic thermal polar cap and nebular fields generally reduce the net IC yield and, being deep in the Klein-Nishina regime at $\gamma_{\rm L}>10^7$, further tighten the energy budget. Absent a population-level closure that maps these assumptions onto $(\dot{\mathcal E},B_\star,P,\alpha)$ and reproduces the FP scalings and the observed $\epsilon_{\rm c}$-$L_\gamma$ trends without fine-tuning, this “ongoing-acceleration synchrotron” narrative remains less economical than the curvature-ECS framework, which attains $\gamma_{\rm L} \gtrsim 10^7$ \emph{in situ} (RRL regime), unifies the GeV and TeV components without extra wind energization, and yields direct, testable population predictions (FP exponents, $\dot{\mathcal E}_{\rm TR}$, and the $P$-$\dot P$ visibility mapping).

A related, complementary line of work models pulsar high-energy emission with an explicit synchro--curvature radiative prescription coupled to an effective (parametrized) description of particle trajectories and emission-region geometry \citep{2024MNRAS.530.1550I,2025MNRAS.541..806I}. In these models the emitting region is assumed to lie just outside the light cylinder, close to the Y-point where the force-free condition breaks and acceleration can occur. The magnetic-field strength and curvature radius sampled by the radiating particles are treated through simple parameterizations along the trajectory rather than being taken from a global FF/PIC magnetosphere \citep{2025MNRAS.541..806I}. As the particle pitch angle is rapidly damped, the emission naturally evolves toward a curvature-dominated gamma-ray component at $\gamma_{\rm L}\gtrsim 10^{7}$ \citep{2024MNRAS.530.1550I}, which is broadly consistent with the central role for curvature emission emphasized in our ECS-based framework. At the same time, because the geometry and acceleration are encoded through effective parameters, establishing a one-to-one mapping between those parameters and the global FF/PIC scalings across the population remains non-trivial. 

\subsection{Open Questions and Future Directions}\label{Open Questions and Future Directions}

The consistency between our theoretical FP scalings, the RRLmax-to-PDL transition, the $P$-$\dot P$ visibility mapping, and the PIC model insights (including a unified account of the GeV curvature and the TeV IC components) is not the whole story. The combined theoretical and observational maturity of this paradigm now lets us pose further incisive questions, aimed at the (unresolved) microphysics that links $\epsilon_{\rm c}$, $L_\gamma$, and the GeV-TeV connection rather than phase-averaged population trends. We highlight three priorities.

\begin{enumerate}[label=(\alph*)]
\item \textbf{Electron-Positron yield versus $P, B_\star$, and $\dot{\mathcal E}$.}
Quantify the magnetospheric pair-production efficiency $\eta_{\rm pair}(\dot{\mathcal E},B_\star,P,\alpha)$ that also is responsible for the observed onset of departures from the RRLmax trend. In practice, $\eta_{\rm pair}$ sets the screening of $E_{\rm acc}$ and thus the effective accelerating scale $\eta_{\BLC}$, fixing the slope change of $\ec(\dot{\mathcal E})$ across MSPs, YPs, and magnetars (and locating $\dot{\mathcal E}_{\rm TR}$). A parallel priority is to assess how multipolar magnetic structure, especially in MSPs, where recent studies suggest significant non-dipolar components \citep[e.g.,][]{2019ApJ...887L..24M,2019ApJ...887L..21R,2020ApJ...893L..38C,2021ApJ...907...63K,2025ApJ...991..169O}, modulates $\eta_{\rm pair}$ and the resulting screening.

\item \textbf{PIC realism and emergent dissipation.}
Global PIC is still under-resolving key scales and processes. To capture the true radiative behavior in a global magnetosphere, simulations must reach realistic magnetic fields, particle energies, and radiation-reaction strengths. Promising avenues include improved particle-orbit integrators \citep[e.g.,][]{2022A&A...666A...5P}, hybrid workflows that couple global MHD/FFE to embedded local PIC \citep[e.g.,][]{2024A&A...690A.170S}, and, potentially, machine-learning surrogates that bridge current physical gaps (e.g., subgrid closures for pair yield and $E_{\rm acc}$ screening in global runs; super-resolution of magnetospheric structure from lower-resolution simulations; fast emulators that extrapolate particle and photon energy distributions to realistic $B_\star$, $\RLC/r_\star$) imposed by numerical limitations.

\item \textbf{Spectral shapes and multi-band couplings.}
Beyond constraining cutoff energies, we need predictive, phase-resolved modeling of the \textit{multiwavelength} pulsed spectrum that explains the richer \citetalias{Smith23} spectral features and, at lower energies, the components that furnish the MeV-IR seed fields for IC. The goal is a unified framework that links the GeV curvature cutoff shape, the MeV band, and the TeV IC tail across viewing geometries, and that also predicts phase-resolved polarization signatures and time variability of the gamma-ray emission. Such a framework would tie $(\ec,L_\gamma)$ to $\dot{\mathcal E}$ in all regimes.
\end{enumerate}

Looking ahead, the framework developed here provides a quantitative bridge between theory and survey design for both space- and ground-based instruments. By linking the FP scalings, the RRLmax-to-PDL transition, the observed deviation from the RRLmax regime, and the $P$-$\dot P$ visibility mapping to observable pairs $(F_\gamma,\epsilon_{\rm c})$, it defines target regions for next-generation MeV-band observatories and VHE $\gamma$-ray facilities, where phase-resolved TeV spectroscopy can test the predicted IC components. In parallel, a companion population-synthesis study builds on these scalings to model Galactic pulsar ensembles with realistic beaming/distance distributions, propagate them through instrument response functions for prospective surveys, and outline observing strategies and timing prospects, including applications to gamma-ray pulsar timing arrays \citep{gammaPTA}. The present work supplies the physical priors, regime structure, and visibility criteria that anchor that broader effort to detect nHz gravitational waves.

\begin{acknowledgments}
We thank the anonymous referee for a careful reading of the manuscript and for constructive suggestions that improved the clarity and presentation of this work. We thank Alexander Philippov, Beno\^{\i}t Cerutti, Abhishek Desai, Christo Venter, George Younes, and Ioannis Contopoulos for helpful discussions. We are also grateful to Liz Hays, Matthew Kerr, and Regina Caputo for their careful reading of the manuscript, valuable feedback, and encouragement. This material is based upon work supported by NASA under awards 80GSFC21M0002 and 80GSFC24M0006, and under grants 21-ATP21-0116, 22-ADAP22-0142, and 22-TCAN22-0027. This work is also supported by the Fermi and NICER missions. Resources supporting this work were provided by the NASA High-End Computing (HEC) Program through the NASA Advanced Supercomputing (NAS) Division at Ames Research Center. This research has made use of the NASA Astrophysics Data System. 
\end{acknowledgments}

\software{}
For the analysis and visualization, Mathematica \citep{Mathematica} is used.

\bibliography{references}{}
\bibliographystyle{aasjournalv7}
\end{document}